\documentclass[11pt,a4paper]{article}
\pdfoutput=1
\usepackage{a4wide}
\usepackage{amsmath}
\usepackage{amssymb}
\usepackage{amsfonts}
\usepackage{cite}
\usepackage[table]{xcolor}
\usepackage{adjustbox}
\usepackage{multirow}
\usepackage{pdflscape}
\usepackage{gensymb}
\usepackage{graphicx}
\usepackage{diagbox}
\usepackage{pifont}
\usepackage{bigstrut}
\usepackage[small,bf]{caption}
\usepackage{enumerate}
\usepackage{tabulary}
\usepackage{float}
\usepackage{arydshln}
\usepackage{enumitem}
\usepackage{mathtools}
\usepackage[utf8]{inputenc}
\usepackage[T1]{fontenc}
\usepackage{footnote}
\usepackage{footmisc}
\usepackage{hyperref}
\usepackage{ctable}
\usepackage{cancel}
 \usepackage[bb=boondox]{mathalfa}
\usepackage{amsfonts}
\usepackage{amsmath,mathrsfs}
\usepackage{amssymb}
\usepackage{mathtools}
\usepackage{color}
\usepackage{graphicx}
\usepackage{epsfig}
\usepackage{mathtools}
\usepackage{ulem}
\usepackage{nameref}
\usepackage{bm}
\newcounter{savefootnote}
\newcounter{symfootnote}
\newcommand{\symfootnote}[1]{%
   \setcounter{savefootnote}{\value{footnote}}%
   \setcounter{footnote}{\value{symfootnote}}%
   \ifnum\value{footnote}>8\setcounter{footnote}{0}\fi%
   \let\oldthefootnote=\thefootnote%
   \renewcommand{\thefootnote}{\fnsymbol{footnote}}%
   \footnote{#1}%
   \let\thefootnote=\oldthefootnote%
   \setcounter{symfootnote}{\value{footnote}}%
   \setcounter{footnote}{\value{savefootnote}}%
}

\newcommand{\sym}{$U(1)_{B-L}$}

\begin{document}

\begin{titlepage}

\vspace*{-15mm}


\centering{\huge{\bf Leptogenesis in an extended seesaw model with $U(1)_{B-L}$ symmetry}}\\[10mm]

\centering
{\bf Ujjal Kumar Dey}$^{a}$\footnote{E-mail: \textit{ujjal@iiserbpr.ac.in}} 
{\bf Tapoja Jha}$^{b,}$\footnote{E-mail: \textit{tapoja.phy@gmail.com}} 
{\bf Ananya Mukherjee}$^{c,}$\footnote{E-mail: \textit{ananya@prl.res.in}} 
{\bf Nirakar Sahoo}$^{\,a,d}$\footnote{E-mail: \textit{nirakar.pintu.sahoo@gmail.com}}\\

\vspace{3mm}
$^{a}$\,{\it Department of Physical Sciences, Indian Institute of Science Education and Research Berhampur, Transit Campus, Government ITI, Berhampur 760010, Odisha, India} \\
\vspace{3mm}
$^{b}$\,{\it School of Physical Sciences, Indian Association for the Cultivation of Science, \\ \it 2A \& 2B Raja S.C. Mullick Road, Kolkata-700 032, India} \\
\vspace{3mm}
$^{c}$\,{\it Theoretical Physics Division, Physical Research Laboratory, Ahmedabad 380009, India} \\
\vspace{3mm}
$^{d}$\,{\it Center of Excellence in High Energy and Condensed Matter Physics, Department of Physics, Utkal University, Bhubaneswar 751004, India} 
\vspace{8mm}

\begin{abstract} 
We have explored an extended seesaw model accommodating a keV sterile neutrino adopting $U(1)_{B-L}$ symmetry. This model provides a natural platform for achieving resonant leptogenesis to account for the observed baryon asymmetry of the Universe. The required lepton asymmetry is sourced by the CP violating decay of the lightest heavy right handed neutrino to Standard Model leptons and Higgs. The presence of the light sterile neutrino in the model brings out an enhancement in the final lepton asymmetry through an additional self-energy contribution. Adopting a proper treatment for all the washout processes this framework strictly favors a strong washout regime thereby protecting the low energy neutrino mass parameters in agreement with the present neutrino and cosmology data. This framework of extended seesaw scheme offers the source of matter-antimatter asymmetry without any severe fine tuning of the Yukawa couplings governing the tiny neutrino mass. We also comment on the half-life period for the neutrino less double beta decay process in the background of having a keV sterile neutrino satisfying all the constraints which guide the explanation for the observed baryon asymmetry of the Universe.
\end{abstract}
\end{titlepage}

\newpage

\hrule \hrule
\tableofcontents
\vskip 10pt
\hrule \hrule 

\setcounter{footnote}{0}
\section{Introduction}
After the youngest discovery of the Standard Model (SM) predictions in terms of the Higgs boson, it is needless to say that the SM is one of the most successful theories of particle physics dealing with the matter and their fundamental interacting forces. In spite of this huge success, there remain a number of limitations of the SM in the context of explaining the origin of neutrino mass, existence of dark matter (DM) and the reason behind the predominance of matter over antimatter in our observable Universe. In the past decades, there have been plenty of theoretical studies with an objective of addressing the above mentioned issues at the cost of constructing some beyond standard model (BSM) frameworks~\cite{Okada:2012fs,Iso:2010mv}. In order to account for tiny neutrino mass one of the most economical possibilities is to embed a seesaw mechanism~\cite{Minkowski:1977sc,GellMann:1980vs,Yanagida:1979as,Mohapatra:1979ia} into the SM. Over the years a host of seesaw scenarios are proposed. The various seesaw models can also have phenomenological importance in the context of probing them in the colliders. Among the variants of the seesaw schemes low-scale seesaw mechanisms have gained considerable attention in view of their testability in the colliders~\cite{Bambhaniya:2016rbb,Hambye:2016sby}. Within the same seesaw framework testing leptogenesis is another probe to validate the neutrino mass generation mechanism from the perspective of various cosmological observations.
%

%
The matter-antimatter asymmetry can be parametrised by the baryon to photon ratio.  Various cosmological observations report this ratio to be~\cite{Planck:2018vyg}
\begin{equation}
\eta_B =\frac{n_B - n_{\bar B}}{n_\gamma} = (6.04 - 6.2)  \times 10^{-10}
\end{equation}
where $n_B$, $n_{\bar B}$, $n_{\gamma}$ and $s$ represent number densities of baryons, antibaryons, photons, and entropy respectively.  A very appealing and field theoretically consistent mechanism of generating the baryon asymmetry of the Universe is the process of leptogenesis which was first pointed out by Fukugita and Yanagida~\cite{Fukugita:1986hr}. Leptogenesis, being a direct consequence of the seesaw models having right handed neutrinos (RHN) also establishes a connection between the origin of neutrino mass and the baryon asymmetry of the Universe. Realization of baryogenesis though leptogenesis via the decay of RHN in various seesaw models can be found in~\cite{Parida:2016asc,Borah:2017qdu,Bambhaniya:2016rbb,Ibarra:2010xw,Granelli:2020ysj,Biswas:2018sib,Croon:2019dfw,Dror:2019syi,Mishra:2019gsr,Rahat:2020mio,Abdallah:2012nm,Biswas:2017tce}. For a recent review one may look at \cite{Davidson:2008bu,Pilaftsis:2009pk,Xing:2020ald,Bodeker:2020ghk} and the references there in. These RHNs can in principle be uncharged under the SM gauge interactions; unless protected by some new symmetry they would mix with the active neutrinos. Inclusion of a pair of sterile neutrinos to the SM fermion sector has become a common lore in order to explain the tiny neutrino mass through type-I seesaw mechanism~\cite{Rubakov:1996vz,Davidson:2008bu,Buchmuller:2004tu,Plumacher:1996kc,Giudice:2003jh}. Type-I seesaw model predicts a bound on the sterile neutrino mass to be very heavy ($M_N \sim10^{14}$ GeV) in order to generate the correct light neutrino masses. However, considering sterile neutrinos at accessible energies gained considerable attention in the last decade from the perspective of being probed directly at high energy collider LHC. Interestingly, the most appealing window for sterile neutrino mass has been seen to be from 1 eV to 10 TeV which also has theoretical and experimental incentives. 
For example, using $\sqrt{s}$ = 13 TeV and 36.1 ${\rm fb}^{-1}$ LHC data the ATLAS~\cite{ATLAS:2018dcj} collaboration performs a search on right-handed heavy $W$ boson ($W_{R}$) and heavy Dirac or Majorana neutrino with the final states of $lljj$ ($l=e,\mu$) with same or opposite sign dilepton. Non-observation of the signals set exclusion limit on $M_{N}$ up to 2.9 TeV in the electron channel and 3.1 TeV in the muon channel, where $W_{R}$ mass ($M_{W_R}$) is 4.3 TeV. Refs.~\cite{Fuks:2020att, Han:2022qgg} study the sensitivity of heavy Majorana neutrino through same sign $W^{\pm}W^{\pm}$ at $\sqrt{s}$ = 13 TeV LHC and also with high luminosity LHC. Opting for benchmark scenario~\cite{Fuks:2020att} adopts a type-I seesaw model and searches for $lljj$ (same sign $l$) final states. Using extensive scanning it has been shown that heavy $M_{N}$ can be probed with the realistic tuning of active-sterile mixing ($\theta_{l4}$). At 95\% C.L. with 300 ${\rm fb}^{-1}$ luminosity $M_{N}$ as heavy as 1-10 TeV can be probed by setting the $\theta_{l4}$ as 0.06-0.3, whereas 20 TeV $M_{N}$ can be probed with $\theta_{l4}$ = 0.5. With 3000 ${\rm fb}^{-1}$ data, same mass values can be probed at 13 TeV LHC with $\theta_{l4}$ as half of the previous values. Similar studies have been done with tri-leptonic final states and missing transverse energy~\cite{Pascoli:2018rsg, Pascoli:2018heg, Han:2022qgg}. 
In~\cite{Pascoli:2018heg} it is shown that $M_{N}$ less than 3.5 TeV can be probed with 15 ${\rm ab}^{-1}$ data at $\sqrt{s}$ = 27 TeV LHC with $\theta_{l4}$ = $10^{-2}$. This article further shows that  100 TeV $pp$ collider can probe heavy Majorana states with $M_{N} <  15$ TeV for $\theta_{l4} < 10^{-2}$. In Ref.~\cite{Ng:2015hba} two scenarios are considered, namely, type-I seesaw and left-right symmetric (LRS) model. The LRS model at 13 TeV LHC with 3000 ${\rm fb}^{-1}$ data can reach $M_{N}$ up to 16 TeV in the region where $M_{N}=M_{W_R}/2$. For type-I seesaw, 100 TeV $pp$ collider with 3 ${\rm ab}^{-1}$ data is able to probe coupling allowed by perturbative unitarity with $M_{N}$ up to 10 TeV for same sign electron channel and up to 5 TeV for same sign muon channel.
In the context of realizing a low scale leptogenesis a mechanism called resonant-leptogenesis ~\cite{Pilaftsis:2003gt, Hambye:2001eu, Hambye:2004jf} has become popular in the past decade. This can be perceived as follows: due to the presence of two quasi-degenerate RH neutrinos, the CP asymmetry parameter which quantifies the leptogenesis process gets resonantly enhanced and opens up possibilities for successful leptogenesis even at RH mass scale $\mathcal{O}$(TeV).
In this work we utilise a $U(1)_ {\rm B-L}$ extension of the SM based on extended seesaw mechanism which offers explanation for the tiny neutrino mass. The features of extended seesaw mechanism can be realized in this way. Apart from offering the tiny neutrino mass, the extended seesaw model also provides a potentially accessible scale for the RHN mass. The low scale RHN being a feature of this seesaw model serves as an excellent candidate which can explain the source of matter-antimatter asymmetry of the Universe though the process of leptogenesis. There have been plenty of studies  which dealt with such low scale leptogenesis from a neutrino mass model embracing the \sym ~symmetry~\cite{Borah:2017qdu,Biswas:2018sib,Chauhan:2021xus,Borah:2021mri}. It is worth mentioning that a low scale seesaw model as embedded in the present framework can also render the gravitino bound harmless~\cite{PhysRevD.71.083502}. Basically in the supersymmetric version of the inverse seesaw, the so-called ``soft-leptogenesis'' scenario, gravitino overproduction sets an upper bound on the reheat temperature which sets constraints on the RHN masses~\cite{Garayoa:2006xs, Garbrecht:2013iga}. In addition the extra neutral fermion singlet in this extended seesaw framework facilitates the process of leptogenesis by its self-energy contribution to the final lepton asymmetry to account for the observed baryon asymmetry of the Universe (BAU). This additional self-energy contribution offered by the presence of the extra fermion singlet ($S$) brings an enhancement in the lepton asymmetry which successfully accounts for the observed BAU. We also emphasize on the correlation among the CP violating phases in the neutrino sector with the leptogenesis parameter space. 

In another front, the observation of neutrino-less double beta ($0\nu\beta\beta$) decay would prove lepton number violation and the Majorana nature of neutrinos~\cite{PhysRevD.22.2227}, a proper investigation of the said process would be of fundamental importance. Here, we attempt to provide the parameter space in the purview of extended seesaw scenario which is compatible with the constraints related to the $0\nu\beta\beta$ decay experiments as reported by KamLAND-Zen \cite{KamLAND-Zen:2016pfg}, and GERDA~\cite{Agostini:2018tnm}. Owing to the presence of the neutral fermion singlets we compute the half-life for the potential $0\nu\beta\beta$ decay process ($T^{0\nu}_{1/2}$). The contribution of the active-sterile mixing to the total $T^{0\nu}_{1/2}$ of this process has been shown. 
In this set-up we present an in-depth analysis on the construction of the high energy neutrino mass matrix in the light of an extended seesaw scheme.  Although the idea of examining an extended seesaw scheme to investigate for the observed BAU  has been conceived before~\cite{Kang:2006sn}, some distinct implications and refinements have not been emphasized earlier. Apart from investigating this seesaw model for resonant leptogenesis, we have also obtained and shown novel findings in the neutrino phenomenology sector which have not been discussed in~\cite{Kang:2006sn} or any other framework based on this extended seesaw scenario. Among these results most important ones are regarding the $\mu -\tau$ symmetry breaking in the lepton sector which has been achieved without considering any additional flavor symmetry and the preference for the higher octant of atmospheric mixing angle $\theta_{23}$. The true octant of $\theta_{23}$ is one of the long-standing puzzles in the study of neutrino oscillation physics. On the other hand, in the leptogenesis sector we have worked in a purely resonant spectrum for the right handed neutrino masses, where the near-degeneracy among the RHN states is the most important requirement for having a successful leptogenesis rather relying on tuning some of the entries of the Yukawa coupling matrix on which ~\cite{Kang:2006sn} is based.
We present the viable parameter space validating this model with respect to the constraints associated with the neutrino oscillation parameters, $0\nu\beta\beta$ process and the leptogenesis. In this work we attempt to focus on a detailed analysis on the high energy neutrino mass parameters associated with the said seesaw model along with a low scale leptogenesis in order to offer an explanation for the observed BAU. Most importantly we have shown that, the present framework allows the required amount of lepton asymmetry to put phenomenologically interesting predictions on the neutrino mixing angles mainly the reactor and atmospheric ones, pushing them towards their higher allowed $3\sigma$ ranges. This fact reinforces the notable features of this analysis. We carried out a detailed analysis to emphasize the interplay between the generated amount of lepton asymmetry and the corresponding washout. 

The outline of this article is as follows. We provide the structure of the model in Section \ref{sec:model}, where we brief the light neutrino mass generation through the extended seesaw mechanism. The necessary simulation details required to extract the ESS parameter space along with their numerical estimation are provided in Section \ref{sec:nu_analysis}.  Section \ref{lepto} is dedicated to the necessary prescriptions required for realizing baryogenesis through leptogenesis in the light of the present set-up.  We discuss the viable parameter space explaining all the above mentioned BSM issues in Section \ref{results}. The constraints associated with $0\nu\beta\beta$ process are discussed in Section \ref{ndbd}. Finally we conclude in Section \ref{conclusion}. Additionally for a detailed description of the construction of the extended seesaw mass matrices the analytical algorithm using the seesaw approximation is shown in \nameref{sec:appenA}. 

\section{Model description} 
\label{sec:model}
In this section, we provide a detailed discussion of the model along with theoretical and some experimental constraints on model parameters. We also provide a brief discussion on the generation of neutrino mass in the extended seesaw framework.
\subsection{Field content and interactions}
\label{sbsc:particles}
We extend the Standard Model by an extra \sym ~symmetry. The model contains three right handed neutrinos ($N_{R_i}$) with $B-L$ charge of $(-1)$ and three complex singlet sterile neutrinos $S_{L_i}$ with vanishing $B-L$ charge. Note that since the triangle anomalies for both the $U(1)_{B-L}^3$ and $U(1)_{B-L}$-(Gravity)$^2$ diagrams are non-zero, it is well-known that the $U(1)_{B-L}$ gauge symmetry is anomalous with SM-only particle content. These triangle anomalies $\mathcal{A}_1^{\rm SM}\left[U(1)_{B-L}^3\right]$ and $\mathcal{A}_2^{\rm SM}\left[U(1)_{B-L}\text{-(Gravity)}^2\right]$ for the SM fermion content turns out to be $(-3)$. In the presence of extra fermions with appropriate $U(1)_{B-L}$ charges these anomalies can be removed. In our case the job is done by the three right handed neutrinos $N_{R_i}$ with $U(1)_{B-L}$ charges $(-1)$. Note that the left handed fermions $S_{L_i}$ with vanishing $U(1)_{B-L}$ charges will not tamper with this cancellation. Apart from these we add two additional scalars $\chi_1, \chi_2$ which are having $B-L$ charges as 1 and 2 respectively. The two scalars break the \sym ~symmetry and give mass to the neutrinos.  
With these two minimal set of singlet scalars we can actually explain the neutrino mass in an extended seesaw set-up. The relevant fields and their respective quantum numbers under the gauge symmetry are shown in table \ref{tab1}. 
\begin{table} [ht]
\centering
\begin{tabular}{|c|c|c|}
     \hline
     Fields & $SU(2) \times U(1)_Y$  &  \sym \\
     \hline
  $(u_L,d_L)^T$    &  2,1/3       & 1/3 \\
  $u_R$           &   1,4/3      &  1/3 \\
  $d_R$           &   1,$-2/3$     &   1/3 \\
  $(\nu_L,e_L)^T$ &  2,$-1$         &  $-1$   \\
  $e_R$           &  1,$-2$        &  $-1$  \\
     $H$            &  2,1          &  0  \\
     \hline
  $N_{R_i}$       & 1,0           &  $-1$ \\
     $S_{L_i}$       & 1,0           & 0  \\
     $\chi_1$     & 1,0          & 1 \\
     $\chi_2$     & 1,0          & 2 \\
\hline
\end{tabular}
\caption{Fields and their quantum numbers under the gauge symmetry.}
\label{tab1}
\end{table}

The Lagrangian for the model can be written as,
\begin{align}
\label{eq:fullLag}
\mathcal{L} = \mathcal{L}_{\rm SM} + \mathcal{L}_{\rm gauge} + \mathcal{L}_{\rm scalar} + \mathcal{L}_{\rm fermion}.
\end{align}
Firstly, within the $\mathcal{L}_{\rm SM}$ itself the gauge fermion interactions will be augmented by new couplings of the SM fermions with the additional $Z^{\prime}$ gauge boson coming from the gauged \sym  ~symmetry. 
In general, from the symmetry point of view the kinetic mixing between $U(1)_Y$ and $U(1)_{B-L}$ is not prohibited and one can always include the relevant kinetic term $(\epsilon/2)B_{\mu \nu}Z^{\prime \mu \nu}$, where $B_{\mu \nu}$, and $Z^{\prime \mu \nu}$ are the field strength of $U(1)_Y$ and $U(1)_{B-L}$ respectively. Since this is highly constrained from electroweak observation, $\epsilon$ can be taken to be zero in the minimal set-up e.g., in~\cite{Basso:2008iv}. It has been shown that the mixing can be generated at the loop-level~\cite{delAguila:1988jz}. In the scales of our interest the effect of this loop-level generation of $\epsilon$ will remain insignificant~\cite{Coriano:2015sea}.
This makes our model simpler and it differs from e.g.,~\cite{Galison:1983pa, Babu:1997st, Leike:1998wr}.
The $\mathcal{L}_{\rm scalar}$ term takes into account all the scalar (including SM-like Higgs doublet) kinetic and mixing terms. This can be explicitly given as,
\begin{align}
\mathcal{L}_{\rm scalar} =
      \left(\mathcal{D}_\mu H \right)^\dagger \left(\mathcal{D}^\mu H\right) 
      +\left(\mathcal{D}_\mu \chi_{1}\right)^\dagger \left(\mathcal{D}^\mu \chi_{1}\right)
       +\left(\mathcal{D}_\mu \chi_{2}\right)^\dagger \left(\mathcal{D}^\mu \chi_{2}\right)
      +V\left(H,\chi_1,\chi_2 \right),
\end{align}
where the respective covariant derivatives are given by,
\begin{subequations}
\begin{align}
\mathcal{D}_\mu H &= \partial_{\mu} H - i\,g \vec{W}_{\mu }\cdot \frac{\vec{\tau}}{2}\, H  \,-\, i\frac{g^{\prime}}{2}B_{\mu} H\, ,\\
\mathcal{D}_\mu \chi_{1} &=\partial_{\mu} \chi_{1} -  i g_{ BL} \,Z_\mu^\prime \chi_{1}  \, , \\
\mathcal{D}_\mu \chi_{2} &=\partial_{\mu} \chi_{2} - 2 i g_{ BL} \,Z_\mu^\prime \chi_{2}\, ,
\end{align}
\end{subequations}
where $g$, $g^{\prime}$, and $g_{ BL}$ are the $SU(2)_{L}$, $U(1)_{Y}$, and \sym couplings, respectively. The $\vec{W}_{\mu},~B_{\mu}$, and $Z^{\prime}_{\mu}$ are the corresponding gauge fields.
The most general scalar potential can be written as,
\begin{align}
V\left(H,\chi_1,\chi_2 \right)  = & -\mu_H^2 H^\dagger H + \lambda_H (H^\dagger H)^2 - \mu_1^2 \chi_1^\dagger \chi_1 + \lambda_1 (\chi_1^\dagger \chi_1)^2 - \mu_2^2 \chi_2^\dagger \chi_2  \nonumber \\ 
  & + \lambda_2 (\chi_2^\dagger \chi_2)^2 +  \lambda_{H1} (H^\dagger H) (\chi_1^\dagger \chi_1) +  \lambda_{H2} (H^\dagger H) (\chi_2^\dagger \chi_2) \nonumber \\
   &  +  \lambda_{12} (\chi_1^\dagger \chi_1)(\chi_2^\dagger \chi_2) +\left[ \mu_{12} \chi_2^\star \chi_1^2 + {\rm H.c.} \right].
\label{eq:scalar_pot}
\end{align}
Here all the $\mu$'s are real positive numbers except $\mu_{12}$. In order to get a positive mass for $CP$-odd scalar states $\mu_{12}$ has to be negative. The scalars $\chi_1 , \chi_2$ break the  \sym ~symmetry and get vacuum expectation values (vev) in this process.  The $SU(2)_{L}$ doublet scalar $H$ is involved in usual electroweak symmetry breaking. So we have two phases of symmetry breaking here,
\begin{align*}
SU(2)_L \times U(1)_Y \times U(1)_{B-L} \xrightarrow{\langle \chi_1 \rangle ,\langle \chi_2 \rangle} SU(2)_L \times U(1)_Y \xrightarrow{\langle H \rangle} U(1)_{\rm em} \,.
\end{align*}
The fields after the symmetry breaking are written as 
\begin{align}
   \chi_1  = \frac{1}{\sqrt 2} (v_1 + \chi_1^\prime + i \rho_1),~~ 
   \chi_2  =  \frac{1}{\sqrt 2} (v_2 + \chi_2^\prime + i \rho_2),~~
    H  = \frac{1}{\sqrt 2}
    \begin{pmatrix}
     \sqrt{2}\omega^{+} \\
      v_H + h + i\zeta
    \end{pmatrix}
\end{align}
After the symmetry breaking three $CP$-even scalars \{$h_{1}, h_{2}, h_{3}$\} will arise out of the mixing of the states \{$h, \chi^{\prime}_{1}, \chi^{\prime}_{2}$\}. Clearly one of \{$h_{1}, h_{2}, h_{3}$\} can be identified as the SM-like Higgs. The $CP$-odd field $\zeta$ is absorbed by the $SU(2)_{L}$ gauge boson hence is not physical.  The other two $CP$-odd fields $\rho_1 , \rho_2$ become massive after the symmetry breaking and they mix with each other. From the $CP$-odd mass matrix one can confirm that one of the eigenvalues is zero. We can say that the field combination $\rho_1 \sin\beta + \rho_2 \cos\beta$ ($\beta$ being the mixing angle between $\rho_1 , \rho_2$) is eaten up by the gauge field $Z^\prime$ and becomes massive, the other combination $\rho_1 \cos\beta - \rho_2 \sin\beta$ is the only physical $CP$-odd field in our model.

Lastly, $\mathcal{L}_{\rm fermion}$ in Eq.~\eqref{eq:fullLag} contains the information of kinetic and Yukawa terms of the BSM fermions and they can be written explicitly as,
\begin{align}\label{yukawa_lag}
\mathcal{L}_{\rm fermion}^{\rm kin} &= \overline{N_{Ri}} i \gamma^\mu \left(\partial_\mu
      - i\,g_\text{BL} \,Z_\mu^\prime \right) N_{Ri}
      + \overline{S_{Li}} i \gamma^\mu \partial_\mu S_{Li},\\
-\mathcal{L}_{\rm fermion}^{\rm Yuk} &= Y_{D_{ij}} \bar{l}_i \tilde{H} (N_R )_j  + Y_{R_{ij}} \chi_2 (N_R)_i ^T (N_R)_j + \mu_{ij} (S_L ^T)_i (S_L)_j + Y_{S_{ij}} \chi_1^\star \bar{N_R}_i (S_L)_j + {\rm H.c.}\,  
\end{align}
here the indices $i,j$ run from 1-3 and represent the generation indices, and we denote $\tilde{H} \equiv i \sigma_2 H^*$.


\subsection{Neutrino Mass generation through extended seesaw mechanism}\label{numass}
  Owing to the simultaneous presence of both the heavy and small lepton number violating scales $M_R$ and $\mu$ respectively, extended seesaw scenario \cite{Kang:2006sn} is very different from the inverse seesaw \cite{Mohapatra:1986bd,Mohapatra:1986aw,Wyler:1982dd,Witten:1985bz,Hewett:1988xc,Dias:2012xp,Dev:2009aw,Blanchet:2010kw,Ilakovac:1994kj,Deppisch:2004fa,Arina:2008bb,Malinsky:2009df,Hirsch:2009ra} and double seesaw scenario \cite{Ellis:1992nq}. In inverse seesaw mechanism, there is only one small lepton number violating scale $\mu$ and the lepton number is conserved in the $\mu= 0$ limit. One of the significant features of this seesaw model is that the presence of the heavy right handed neutrinos do not directly control the generation of the tiny neutrino masses, while playing a crucial role in yielding relatively light sterile neutrinos. Another important characteristic of this extended seesaw mechanism is that, one can have contribution of Majorana neutrinos to $0\nu \beta \beta$ transition amplitude even in the limit $\mu \rightarrow 0$, whereas in the inverse seesaw scenario, the corresponding amplitude vanishes.  
  
 One can write the complete neutrino mass matrix which is generated after the symmetry breaking as follows. From Eq.~\ref{yukawa_lag} one can derive the neutrino mass matrix as,
 \begin{equation}
   \mathcal{M}_{n}=
   \begin{pmatrix}
     \nu_L & S_L  & N_R^c 
   \end{pmatrix}
   \begin{pmatrix}
     0  &  0  & M_D^T \\
     0  &  \mu  & M_S^T \\
     M_D &  M_S  & M_R
   \end{pmatrix}
   \begin{pmatrix}
     \nu_L^c \\
     S_L^c \\
     N_R 
     \end{pmatrix} \, ,
 \end{equation}

   where the matrix elements are given as:
   \begin{eqnarray}
     M_D &=& Y_{D} v_H \,, \\ \nonumber
     M_S &=& Y_{S} v_1 \,, \\ \nonumber
     M_R &=& Y_R v_2 \,.
   \end{eqnarray}

To the leading order, the light neutrino mass matrix $m_\nu$, and the heavy neutrino mass matrices $m_s$, $m_R$ can be constructed from the following equations,
\begin{flalign}
\label{eq:matrices}
&~~~~~~~~~~~~~~~~~~~~~~~~~~~~~~~~~~~~~~~~~~~~~~m_\nu \sim M_D^T(M_S^T)^{-1}\mu(M_S)^{-1}M_D&\\ 
&~~~~~~~~~~~~~~~~~~~~~~~~~~~~~~~~~~~~~~~~~~~~~~m_s \sim \mu - M_S^T M_R^{-1}M_S&\\
&~~~~~~~~~~~~~~~~~~~~~~~~~~~~~~~~~~~~~~~~~~~~~~m_R \sim M_R&
\end{flalign} 
To this end, we refer the reader to 
~\nameref{sec:appenA} for a detailed construction of the extended seesaw mass matrix following the seesaw parametrization.

\subsection{Constraints on model parameters}
\label{sbsc:constraints}
In this section we briefly discuss some of the theoretical and observational constraints on the model parameters.
The boundedness of the scalar potential puts constraints on the respective quartic couplings. Following the method of co-positivity~\cite{Kannike:2012pe, Chakrabortty:2013mha}, one can easily write down the germane conditions which in this case are,
\begin{subequations}
\begin{gather}
\lambda_H \,, \lambda_1 \,, \lambda_2 >0 \,, \\
\bar{\lambda}_{H1} \equiv \lambda_{H1} + 2 \sqrt {\lambda_H \lambda_1} >0 \,, \\
\bar{\lambda}_{H2} \equiv \lambda_{H2} + 2 \sqrt {\lambda_H \lambda_2} >0 \,, \\
\bar{ \lambda}_{12} \equiv \lambda_{12} + 2 \sqrt {\lambda_1 \lambda_2} >0 \,, \\
\lambda_{H1} \sqrt{\lambda_2} +  \lambda_{12} \sqrt{\lambda_H} +  \lambda_{H2} \sqrt{\lambda_1} + 2 \sqrt{\lambda_H \lambda_1 \lambda_2} + \sqrt{2  \bar{\lambda}_{H1}  \bar{\lambda}_{H2}  \bar{ \lambda}_{12}} > 0 \,.
\end{gather}
\end{subequations}
%
%
The LEP constraint on the ratio of new gauge boson mass to the coupling is \cite{Schael:2013ita,Carena:2004xs}
\begin{equation}
\frac{M_{Z^\prime}}{g_{BL}} \geq 6 \rm TeV.
\end{equation}
Similar bound from ATLAS is quite less stringent $M_{Z^\prime}/g_{BL} \geq 4.1$ TeV~\cite{Aaboud:2017buh}.

In the following section we discuss the methodology used to numerically evaluate the allowed parameter space satisfying the neutrino mass and mixing parameters which comply with the $3\sigma$ global fit presented in table \ref{tabdata}. 

\section{Numerical analysis of extended seesaw parameter space}
\label{sec:nu_analysis}
Before discussing the leptogenesis aspects of the model we first consider the constraints on the parameter space coming from the neutrino oscillation data. In this section we describe the methodology used to extract the ranges of the various parameters that can generate the sub-eV light Majorana neutrino mass within the extended seesaw scheme (ESS). 
As described in the model section the ESS framework is realized by three generations of sterile neutrinos $S_{L}$ along with another three generations of heavy right-handed neutrinos $N_{R}$. Thus the high energy neutrino mass matrix basically turns out to be of the dimension $9 \times 9$. In order to diagonalize this $9 \times 9$ mass matrix analytically we follow the procedure given in Refs.~\cite{Grimus:2000vj, Mitra:2011qr}. A detailed construction of mass matrices and the corresponding diagonalizing matrices are given in ~\nameref{sec:appenA}. We first follow the two steps diagonalization: in the first case we are left with one $9 \times 9$ block-diagonalized matrix comprising of three $3 \times 3$ mass matrices: $m_{\nu}$, $m_{s}$ and $M_{R}$ for active neutrino and two heavy neutrino states, respectively. The expressions of these mass matrices are provided in Sec.~\ref{numass}. The matrices $\mu$, $M_{D}$, $M_{S}$ are considered to be complex symmetric in nature to comply with a non-zero CP phase which can potentially act as a prime source of sizeable lepton asymmetry. However, for simplicity $M_{R}$ is chosen to be a real diagonal matrix. This primary choice of the diagonal $M_{R}$ also ensures that we are working in a basis of Dirac Yukawa coupling matrix where the RHNs are in their physical mass basis. 
%

%
For our numerical analysis we vary each elements of the mass matrices in the following ranges (in GeV):
\begin{subequations}
\begin{align}\label{eq:ranges}
10^{-7} &\le {\mu}^{R}, {\mu}^{I}  \le 10^{-5} \,, \\ 
10^{-5} &\le M_{D}^{R}, M_{D}^{I}  < 10^{-3} \,, \\ 
0.01 &\le M_{S}^{R}, M_{S}^{I} < 0.5,
\end{align}
\end{subequations}

whereas, $M_{{R}_{1}}$ and $M_{{R}_{2}}$ are randomly varied from 10 to 20 TeV and are kept to be nearly degenerate in order to make this framework suitable for resonant leptogenesis. The ratio of these two scales is crucial for leptogenesis in the present scenario which we discuss elaborately in the Sec.~\ref{lepto}. Following the hierarchy of the RHN mass eigenvalues as $M_{{R}_{1}} \approx M_{{R}_{2}} < M_{{R}_{3}}$ we proceed for the numerical diagonalization procedure. We keep $M_{{R}_{3}}$ in 50 TeV to 80 TeV range so that it is completely decoupled from other two right-handed neutrinos. It is worth to note that due to the sub-eV neutrino mass one has to be careful enough while choosing the mass scales associated with the other mass matrices present in a seesaw model. From Eq.~\ref{eq:matrices} we can see the effective neutrino mass scale in our model is $\mu M_D^2/M_S^2$ which should be of the order of $10^{-3} - 10^{-2}$ eV. 
Besides this, we also know that having sterile neutrino in a seesaw model can induce non-unitarity~(NU) on the lepton mixing matrix (also known as the PMNS matrix). 
The NU parameter measures the deviation from the unitarity of the PMNS matrix. The NU arising from the heavy-active mixing is parametrized as \cite{Blennow:2016jkn}, 
\begin{equation}
N = (1- \alpha)U,
\end{equation}
where, $U$ is the usual PMNS matrix, $\alpha$ measures the deviation. In this ESS model the relevant order for $\alpha$ is decided by $\frac{1}{2}M_{D}^{\dagger}{M_{S}^{-1}}^{\dagger}M_{S}^{-1}M_{D}$ (see Eq.~\ref{eq:Ufinal}). For more detail on the NU scale one may look into Ref.~\cite{Blennow:2016jkn}. Taking into account this effect,  one can notice that the ratio $M_D^2 / M_S^2 < 10^{-3,-2}$. So in our analysis we have safely maintained that constraint, which is found to oscillate between $10^{-6} - 10^{-4}$. In order to satisfy the light neutrino mass scale we have chosen
 $\mu$ to be of the order of $10^{-7} - 10^{-5}$~GeV. It is evident that the required scales of the parameter space which essentially satisfy neutrino oscillation data, also satisfied by the NU parameter which comply with the experiment. We also investigate whether this parameter space can explain the leptogenesis. Additionally, we check the viability of the parameter space in the context of neutrino less double beta decay. Therefore, it is quite understandable that any random choice for the ESS parameter spaces can not account for the aforementioned observables.

It is evident that working with the $9 \times 9$ mass matrix with many undetermined parameters may be quite daunting. For our purposes we generate such a large dimensional complete mass matrix numerically for convenience. To simulate this we use \texttt{Python} programming which essentially yields the relevant parameter space keeping the constraints on the various mass scales of such ESS framework intact.
The complete simulation maintaining neutrino oscillation data as well as desired value of the lepton asymmetry has been performed by in-house \texttt{Python~3.8.3} code.
From this simulation we obtained parameter points which essentially yield the neutrino oscillation observables. The obtained neutrino mixing parameters are then compared with the current neutrino data pertinent for normal hierarchy (NH), as summarised in table \ref{tabdata}. We should mention here that, the present framework is suitable to establish both of the possible neutrino mass hierarchies, however for simplicity and to be in agreement with the recent bias for the NH~\cite{Capozzi:2018ubv} we only stick to NH of neutrino mass and carry out the analysis relevant for the same.
\begin{table}
\begin{center}
\begin{tabular}{|c|c|c|}
\hline
Parameters & Normal ordering & Inverted ordering  \\
\hline \hline
$\sin^2\theta_{23}$ & 0.433 - 0.609 & 0.436 - 0.610\\
\hline
$\sin^2\theta_{12}$ & 0.275 - 0.350 & 0.275 - 0.350 \\
\hline
 $\sin^2\theta_{13}$ & 0.02044 - 0.02435 & 0.02064 - 0.02457 \\
\hline
$\Delta m^2_{21}$ & (6.79 - 8.01)$\times 10^{-5}$ eV$^2$ & (6.79 - 8.01 ) $\times 10^{-5}$ eV$^2$\\
\hline
$\Delta m^2_{31}$ & (2.436 - 2.618)$\times 10^{-3}$ eV$^2$ & $-$(2.601 - 2.419)$\times 10^{-3}$ eV$^2$\\
\hline
$\delta_{CP}/ ^{\circ}$ &144 - 357 & 205 - 348 \\
\hline
\end{tabular}
\caption{Latest $3\sigma$ bounds on the oscillation parameters (With SK data) from Ref.~\cite{Esteban:2018azc}.}
\label{tabdata}
\end{center}
\end{table}
At present the value of Dirac CP phase is ambiguous in the sense that, T2K experiment~\cite{Abe:2019vii} prefers $\delta_{\rm CP} \approx \pi/2$  whereas  NOvA~\cite{Acero:2019ksn} tells that $\delta$ can take CP conserving values. However, the global fit values for each hierarchy imply $\delta_{\rm CP} \approx -3\pi/4$ for NH and $\delta_{\rm CP} \approx -\pi/2$ for IH as also evident from the table \ref{tabdata}. Taking care of the above neutrino oscillation constraints we extract the allowed region of parameter space which we discuss in the following subsection. 

\subsection{ESS paramater space}\label{ESS parameter}
\begin{figure*}[t]
\begin{center}
\includegraphics[scale=0.5]{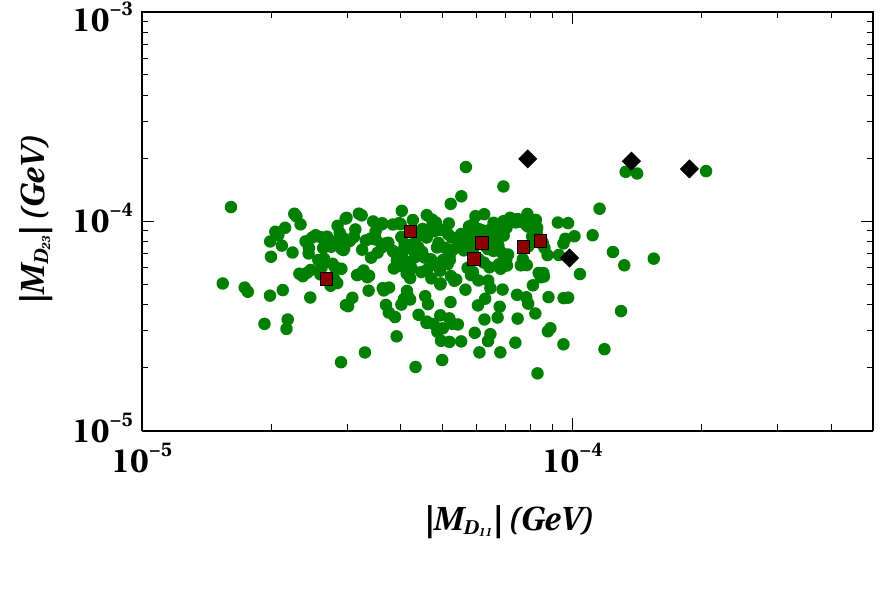}
\includegraphics[scale=0.5]{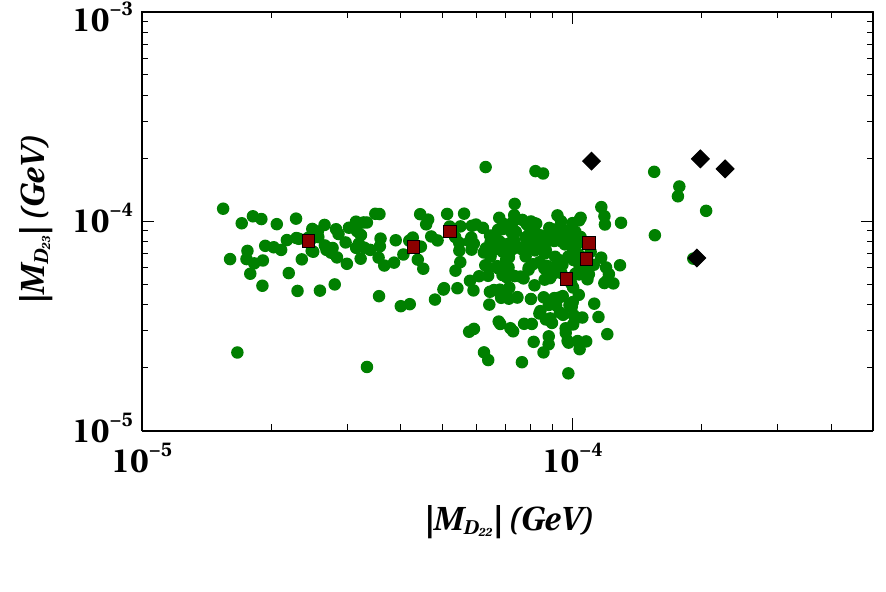} \\ 
\includegraphics[scale=0.5]{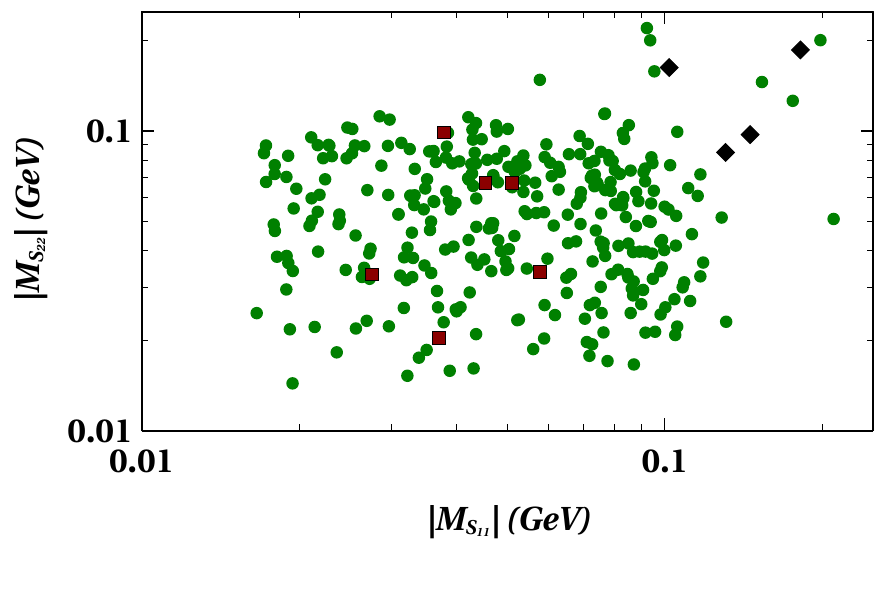}      
\includegraphics[scale=0.5]{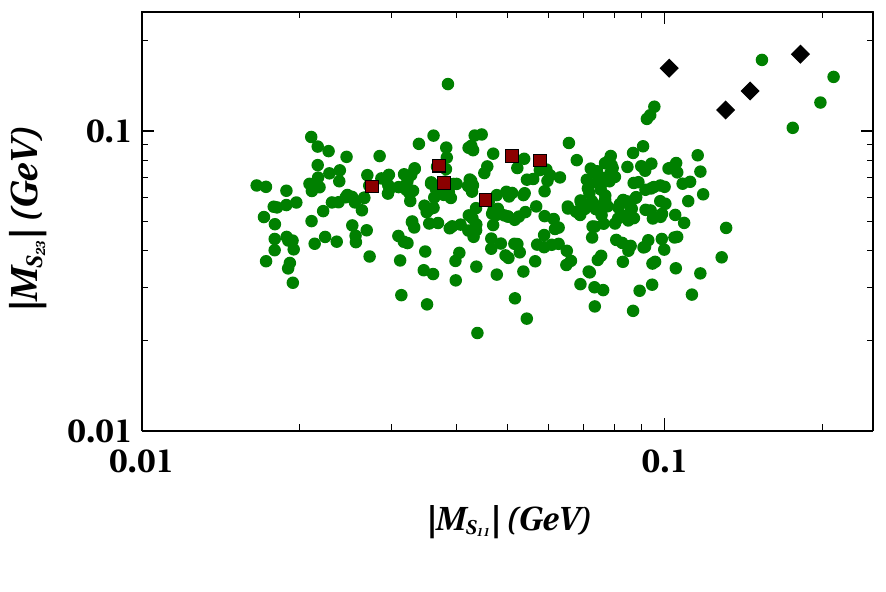}     
\caption{Possible correlations among matrix elements. This also shows the constrained parameter space required to have an adequate amount of lepton asymmetry to account for the observed BAU. The red coloured points are identified with the actual data points which can yield $\eta_B = (2.6 - 6.2)\times 10^{-10}$. Detailed explanations are provided in the text.}
\label{mumd}
\end{center}
\end{figure*}

\begin{figure*}[t]
\begin{center}
\includegraphics[scale=0.5]{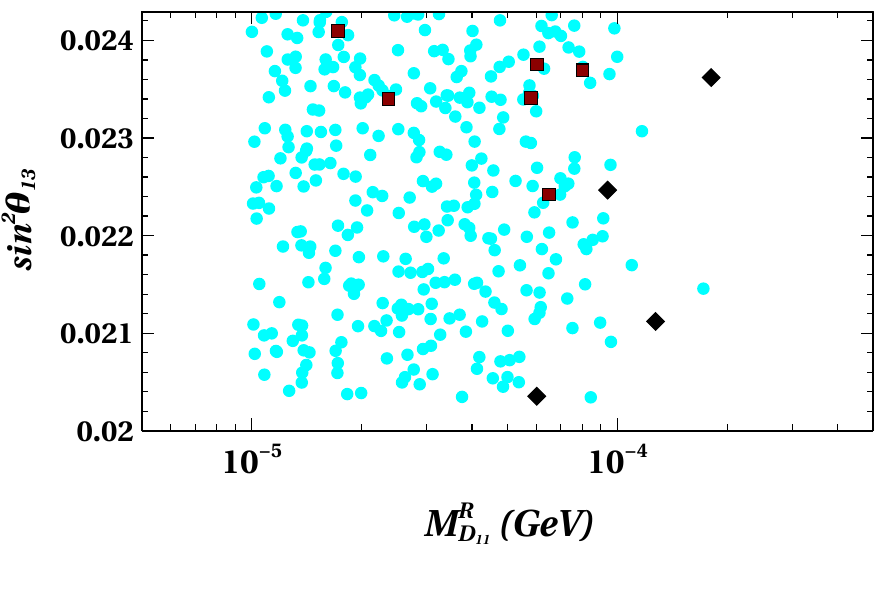}
\includegraphics[scale=0.5]{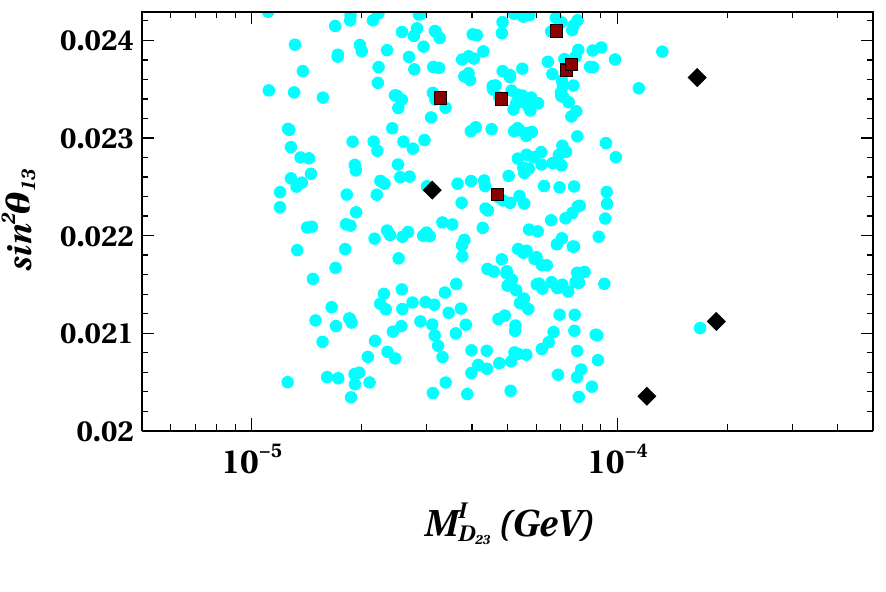} \\ 
\includegraphics[scale=0.5]{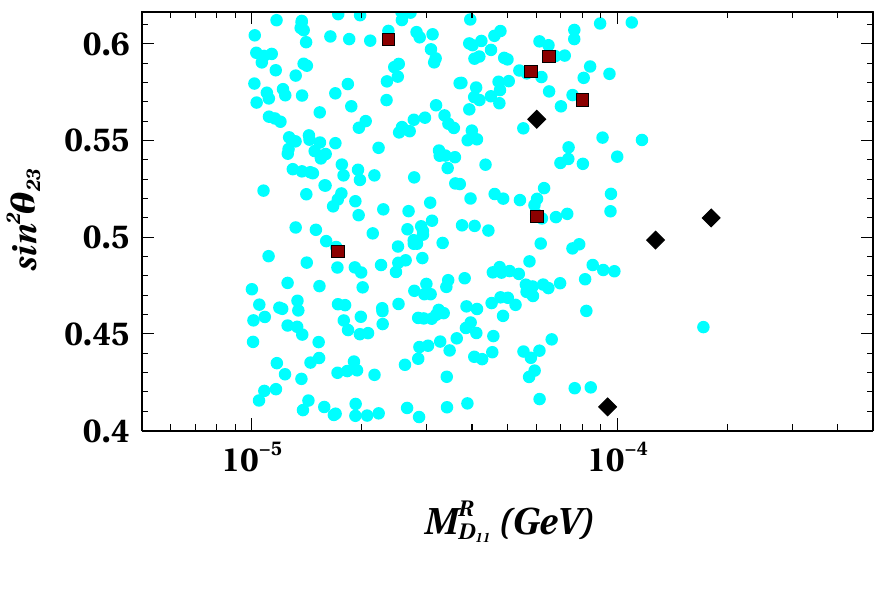}
\includegraphics[scale=0.5]{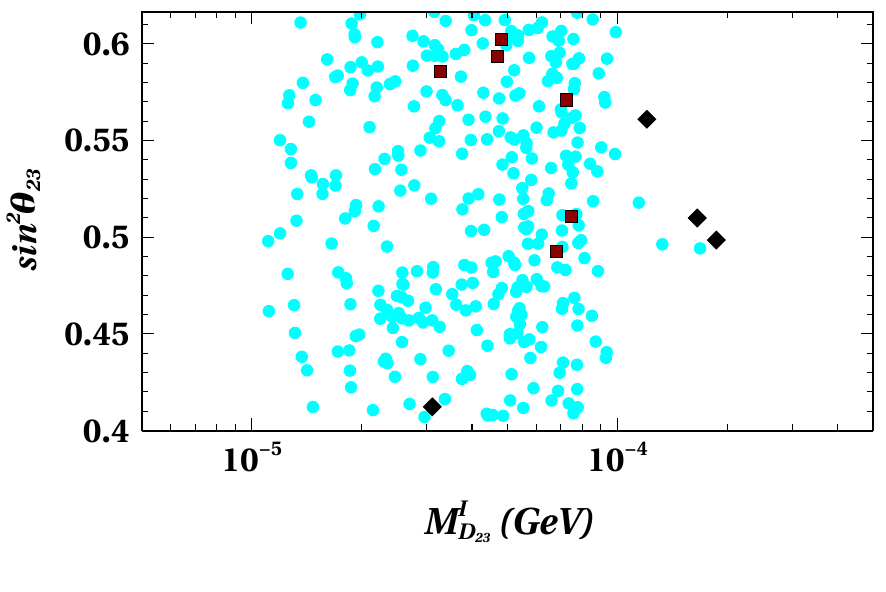}    
\caption{Reactor mixing angle ($\sin^2 \theta_{13}$) (upper panel) and atmospheric mixing angle ($\sin^2 \theta_{23}$) (lower panel) as a function of the ESS mass matrix elements. The red points denote the relevant data which can account for the observed BAU. Detailed explanations are in the text.}
\label{reactor1}
\end{center}
\end{figure*}

\begin{figure*}[h!]
\begin{center}
\includegraphics[scale=0.28]{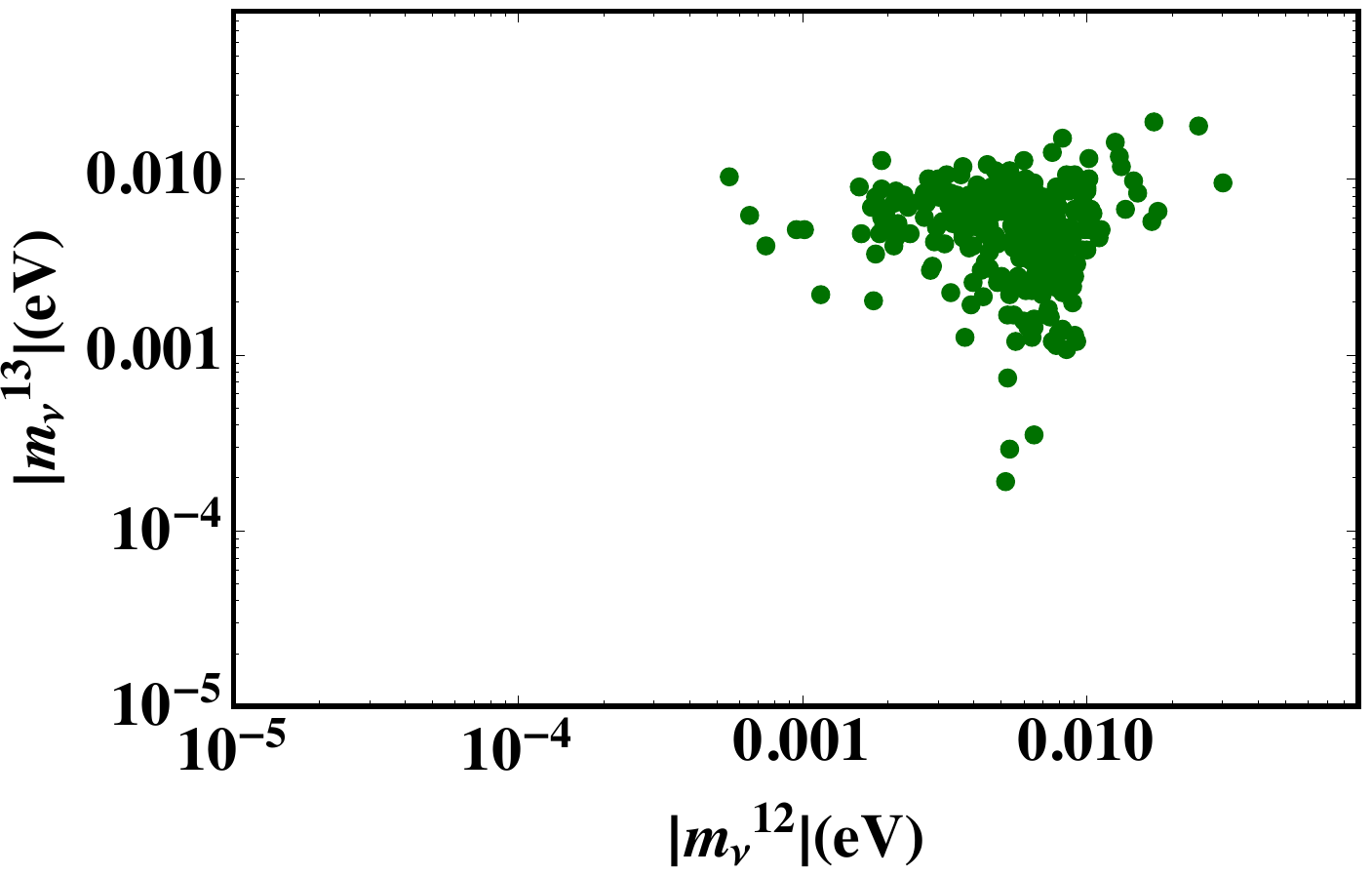}
\includegraphics[scale=0.3]{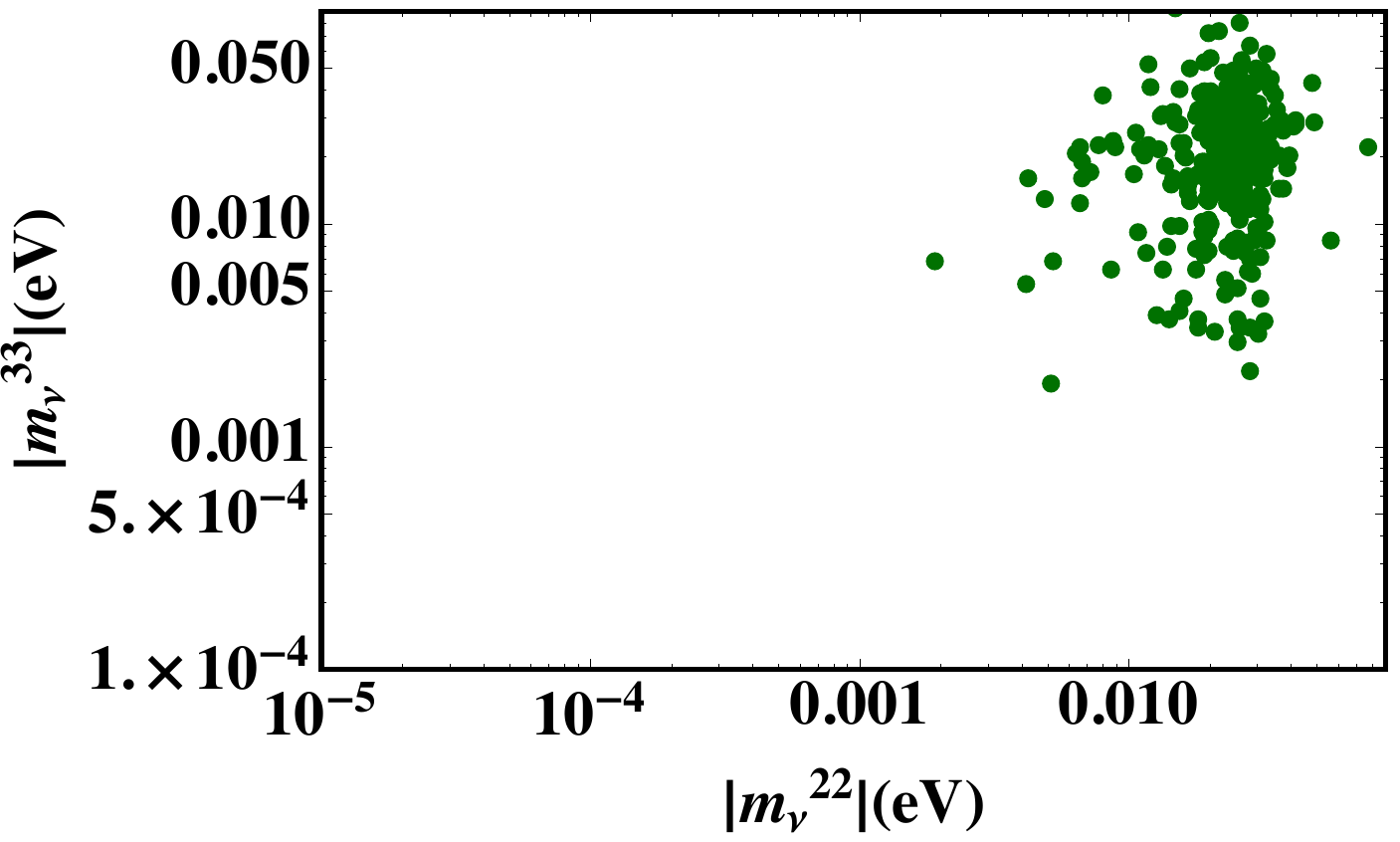}    
\caption{Correlation plots among the low energy neutrino mass matrix elements, showing the plane $m_{e \mu} - m_{e \tau}$ and $m_{\mu \mu} - m_{\tau \tau}$.}
\label{reactor2}
\end{center}
\end{figure*}

\begin{figure*}[h!]
\begin{center}
\includegraphics[scale=0.5]{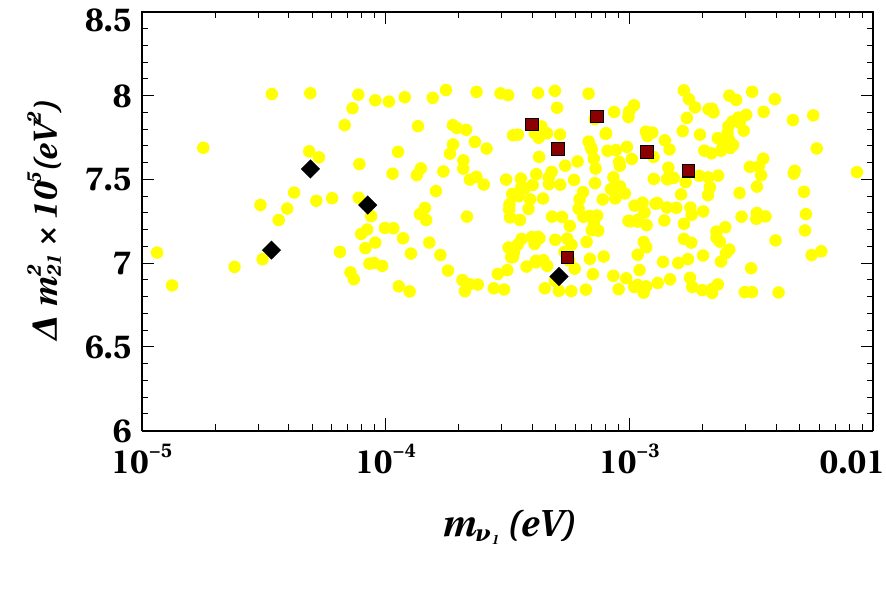}
\includegraphics[scale=0.5]{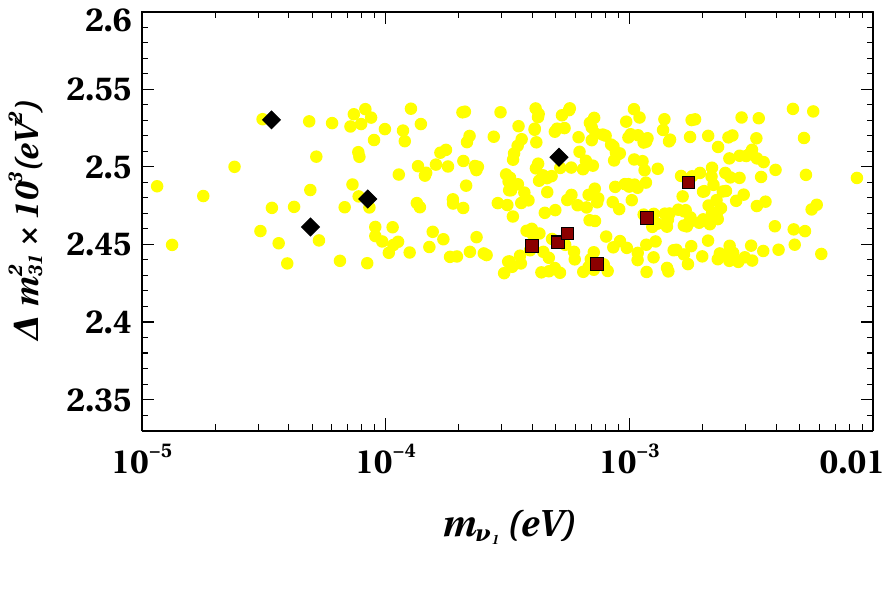} \\    
\caption{Atmospheric and solar neutrino mass squared splittings with respect to the lightest neutrino mass. Red colored squares indicate the relevant data needed to generate an adequate amount of lepton asymmetry. Black colored points are ruled out from the perspective of leptongenesis.}
\label{mass_splitting}
\end{center}
\end{figure*}

\begin{figure*}[h!]
\begin{center}
\includegraphics[scale=0.5]{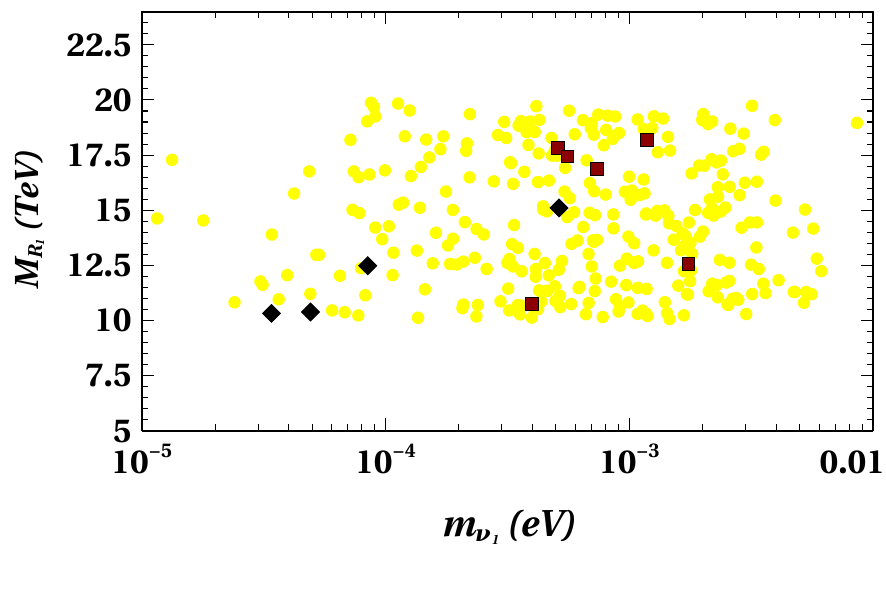}
\includegraphics[scale=0.5]{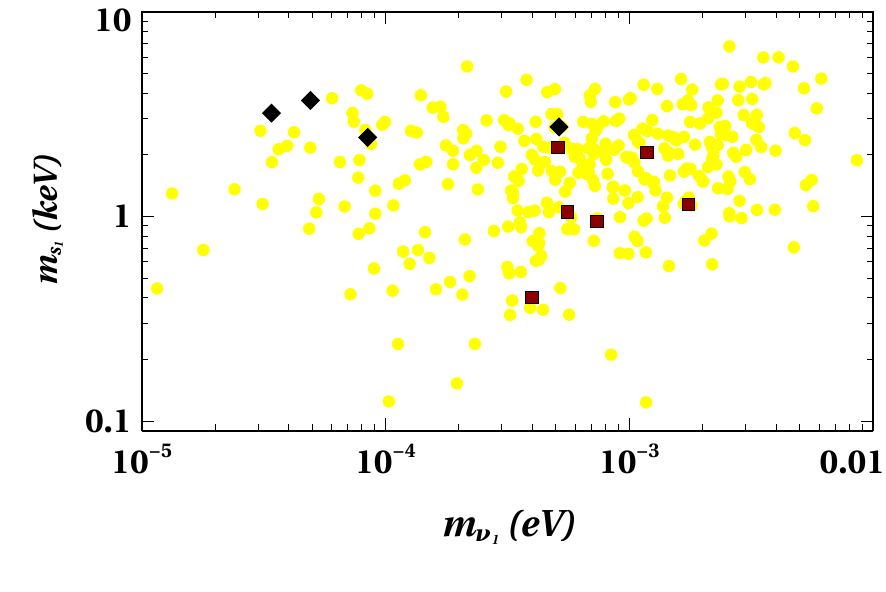} \\    
\caption{Heavy sterile and light sterile masses with respect to lightest Majorana neutrino mass. The choice of the colors can be referred to the same as fig.~\ref{mass_splitting}.}
\label{heavylight}
\end{center}
\end{figure*}

\begin{figure*}[h!]
\begin{center}
\includegraphics[scale=0.5]{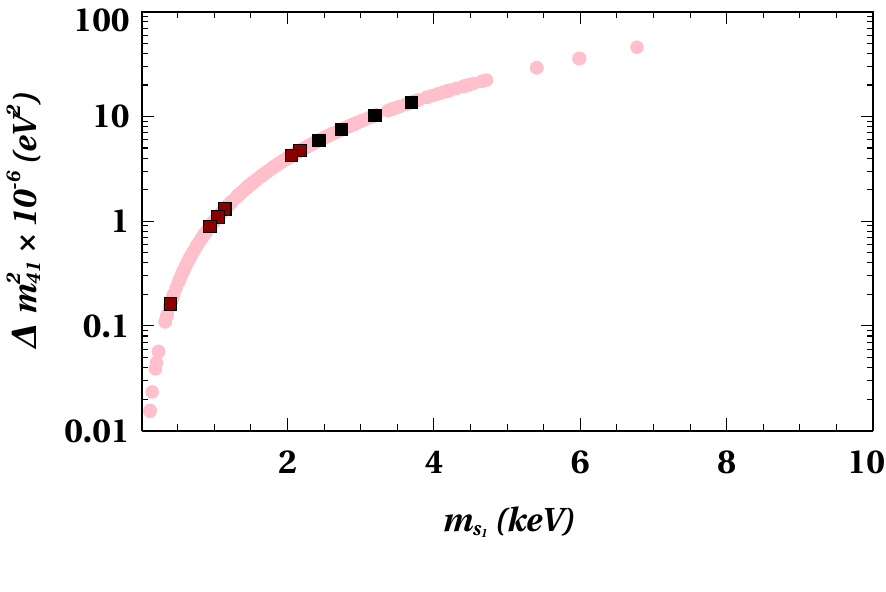}  
\includegraphics[scale=0.5]{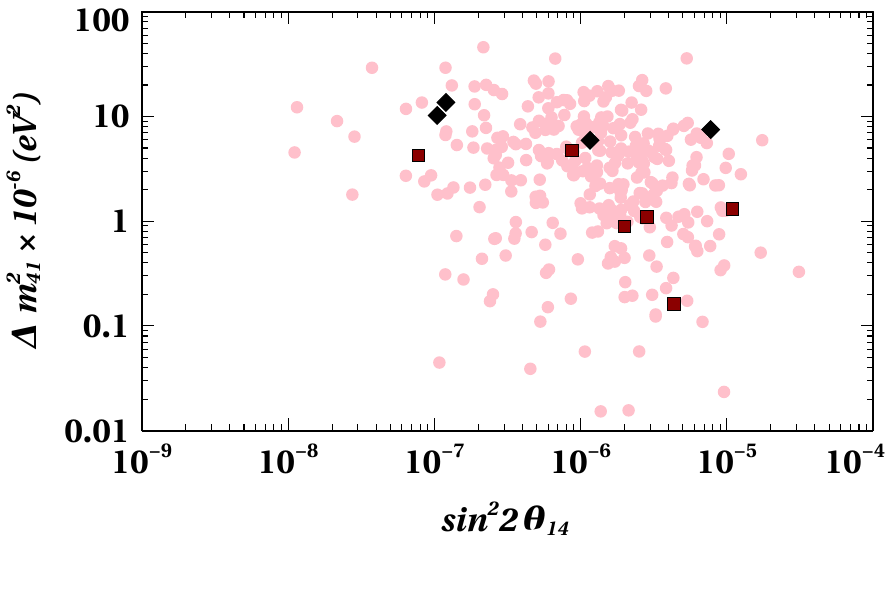} 
\caption{Active-sterile mass squared difference as a function of the lightest sterile neutrino mass (left panel) and $\sin ^2 2 \theta_{14}$ (right panel). The red small squares represent the data which yield ample amount of lepton asymmetry that is responsible for the observed $\eta_B$. The black colored rectangles indicate the data points which are ruled out from the need of sufficient leptogenesis.}
\label{sterilemixing}
\end{center}
\end{figure*}

After constructing the extended seesaw mass matrix numerically ($\mathcal{M}_{n}$) we diagonalise the same. 
From the diagonalizing matrix we extract the values of the mixing angles and the CP phases involved in the lepton mixing matrix~\cite{ANTUSCH2003401}. fig.~\ref{mumd} evinces a mild correlation among the matrix elements responsible to generate the neutrino observables which comply with experimental data.Here we use different color codes to highlight the required parameter values which are useful for successful leptogensis.
In principle correlations among the neutrino mixing parameters would severely constrain the ESS model parameter space, which may make the model phenomenologically richer. This issue can be well-taken care of if one introduces a flavor symmetry embedded framework which leads to a particular texture of the low energy neutrino mass matrix ($m_\nu$) giving rise to possible correlations among the neutrino observables \cite{Borah:2017qdu, Mukherjee:2017pzq, Krishnan:2020xeq}.
In fig.~\ref{reactor1} we present the variation of the reactor ($\theta_{13}$) and atmospheric ($\theta_{23}$) mixing angles with respect to the real and imaginary parts of the matrix element of $M_D$. It is evident from fig.~\ref{reactor1} that the entire range of the ESS parameter space associated with all the matrix elements can yield a non-zero value of $\theta_{13}$ and $\theta_{23}$ compatible with the global fit as mentioned in table ~\ref{tabdata}. A similar notion is obtained for the other matrices ($\mu$ and $M_S$) also, when plotted (not shown here) with respect to these mixing angles of the leptonic mixing matrix. The red points represent the parameter space which can yield sufficient amount of lepton asymmetry leading to the observed $\eta_B = (2.6 - 6.2)\times 10^{-10}$. On the other hand the black points indicate the regions of parameter space which generate ample lepton asymmetry, but can not give rise to the observed $\eta_B$ due to large amount of washouts associated with them. One can notice from these figures that, in view of successful leptogenesis there exist slight preferences for the higher octant~(HO) of the $\theta_{23}$ and higher values of the $\theta_{13}$ in its allowed $3\sigma$ range (shown by the red points). One can understand the breaking of the $\mu-\tau$ symmetry \cite{Rahat:2018sgs,Perez:2019aqq,Fukuyama:2020swd} here by showing the correlations among the mass matrix elements ($m_{{\nu}_{ii}}$, Eq.~\ref{eq:matrices}) of the $\mu - \tau$ sector as presented in fig.~\ref{reactor2}. Generation of the non-zero reactor angle can be realized by such correlations. Mainly to generate a non-zero reactor mixing angle one has to break the $\mu - \tau$ symmetry in the neutrino mass matrix, for which one needs to take correction to the leading order neutrino mass matrix, which can also be driven by the presence of discrete flavor symmetries (see for instance Ref.~\cite{Borah:2017qdu, Mukherjee:2017pzq, Krishnan:2020xeq}). We have not imposed any such discrete flavor symmetry in our model. However, the neutrino mass matrix thus constructed is found to be of complex symmetric nature, reproducing a non-zero reactor angle. 
We show the variation of the solar~($\Delta m^2_{21}$) and atmospheric~($\Delta m^2_{31}$) neutrino mass splittings as a function of the lightest neutrino mass in fig.~\ref{mass_splitting}. Fig.~\ref{heavylight} shows the allowed ranges of the heavy and light sterile neutrinos with respect to the lightest active neutrino mass $m_{\nu_1}$ we obtain in this analysis. It is evident from fig.~\ref{mass_splitting} and fig.~\ref{heavylight} that slightly higher values of the lightest active neutrino mass (denoted by red squares) are preferred, owing to the required leptogenesis parameter space.  In fig.~\ref{sterilemixing} we present the fourth (active-sterile) mass squared difference ($\Delta m_{41}^2$) as a function of the lightest sterile neutrino mass ($m_{s_1}$) and the active-sterile mixing angle obtained in the set-up. From the right panel of the fig.~\ref{sterilemixing} one can notice the smallness of the active-sterile mixing angle ($\sin 2\theta^2_{14}$) which implies that $\theta^{2}$ mainly oscillates below $ 10^{-5}$, providing the desired values of lepton asymmetry with $m_{s_i}\sim \mathcal{O}$(keV).

Having all the above ESS parameter space, allowed by the neutrino oscillation data, we feed them to lepton asymmetry calculation in order to check for the viable parameter space which can yield the expected  baryon to photon ratio ($\eta_B$). The salient features of the various mass scales (and hence in turn the associated Yukawa couplings) in the context of resonantly enhanced lepton asymmetry will be discussed in the following sections.

\section{Baryogenesis through leptogenesis}\label{lepto}
\subsection{Lepton asymmetry parameter}
\begin{figure*}[h!]
\begin{center}
\includegraphics[scale=0.41]{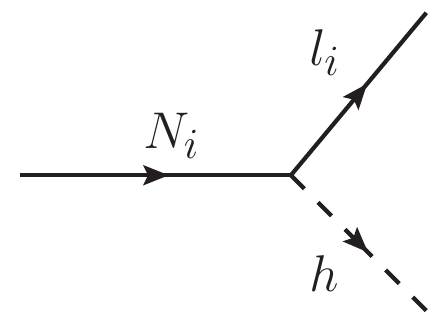}~  
\includegraphics[scale=0.31]{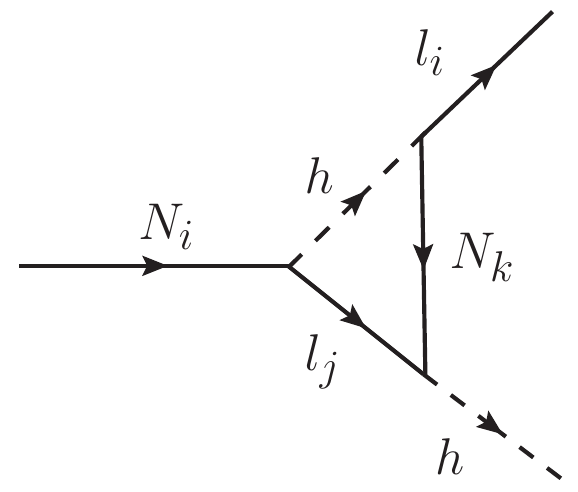}~
\includegraphics[scale=0.41]{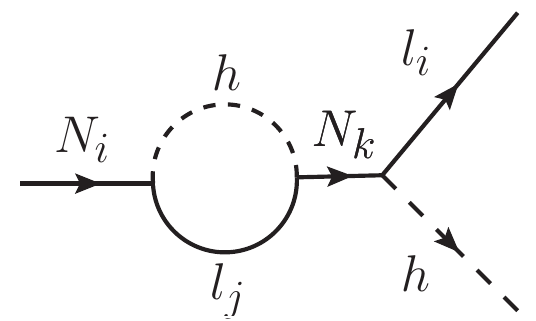}~  
\includegraphics[scale=0.41]{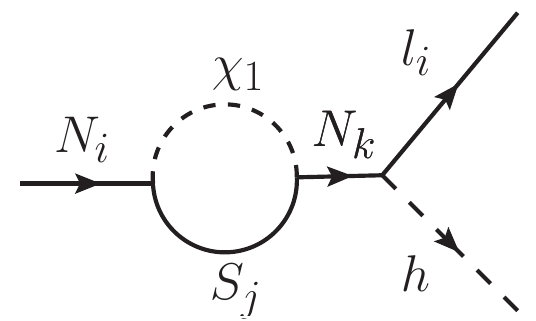}
\caption{Lightest right-handed neutrino decay processes that lead to the lepton asymmetry.}
\label{fig:feynDiag}
\end{center}
\end{figure*}

The lepton number asymmetry sourced by the decay of the RHN can be realized as follows 
\begin{equation*}
\epsilon_i = -\left[ \frac{\Gamma(N_i\rightarrow \bar{l_i}H)-\Gamma(N_i\rightarrow l_i H^*)}{\Gamma(N_i\rightarrow \bar{l_i}H)+\Gamma(N_i\rightarrow l_i H^*)}\right]
\end{equation*}
where, $N_i$ is the lightest RHN creating the lepton asymmetry. 
We present the relevant feynmann diagrams for the decay of the lightest RHN in fig~\ref{fig:feynDiag}. It is important to note here that, in this ESS framework the existence of the fourth self-energy diagram participate significantly in bringing out an enhancement of the final lepton asymmetry. We address this issue in detail in the result section of the lepton asymmetry parameter.
For massless limit of the final state particles of the above decay, the lepton asymmetry can be expressed as  \cite{Kang:2006sn}
\begin{equation} \label{leptonasymmetry}
\epsilon_1 = \frac{1}{8\pi}\sum_{k\neq1}\left(\left[g_{v}(x_k)+g_{s}(x_k)\right]T_{k1}+g_{s}(x_k)S_{k1}\right)
\end{equation}
with $T_{k1} = \frac{\rm Im[(Y_D Y_D^\dagger)^2_{k1}]}{(Y_D Y_D^\dagger +Y_S Y_S^\dagger)_{11}}$ and $S_{k1} = \frac{\rm Im[(Y_D Y_D^\dagger)_{k1} (Y_S^\dagger Y_S)_{1k}]}{(Y_D Y_D^\dagger +Y_S Y_S^\dagger)_{11}}$. 
The Yanagida loop factors are defined as $g_{v}(x)= \sqrt{x_k}\left(1-(1+x)\text{ln}\frac{1+x}{x}\right)$ and $g_{s}(x)= \frac{\sqrt{x_k}}{1-x_k}$ where, $x_k = \frac{M_{R_k}^2}{M_{R_1}^2}$ for $k\neq1$. 
As mentioned earlier, a low scale RHN can overcome the gravitino overproduction problem since it requires a smaller reheating temperature. At the same time for a TeV scale RHN being naturally offered by the seesaw scenario under consideration opens up the window to have the leptogenesis taking place through the resonant enhancement of the leptonic asymmetry \cite{Pilaftsis:2003gt}. Resonant leptogenesis require a very small splitting among the heavy RHN masses of the order of their individual decay width i.e., $M_i - M_j \approx \frac{\Gamma_j}{2}$ \cite{Pilaftsis:2003gt}. In this scenario of large enhancement, the lepton asymmetry can even reach close to unity $\epsilon_1^l \sim 1$. Purposefully,  we have assumed the heavy right handed Majorana neutrinos $N_i$ having a quasi degenerate mass spectrum, following $M_{3} > M_1\left(\approx M_2\right)$ so that any two of the three RHNs can contribute in the leptogenesis process. This leads to the realization that studying the evolution of the number density of $N_1, N_2$ is sufficient in order to obtain their thermal abundance through Boltzmann's equations discussed in the following subsection. 
%
\subsection{Boltzmann Equations }
Solution to the Boltzmann equation \ref{beq} provides us with the dynamics of the RHN production and the $B-L$ number density. From the requirement of producing a lepton asymmetry of the correct order of magnitude, the RHNs have to be numerous before they decay. This essentially relies on the fact that, they are in thermal equilibrium at high temperatures. It is worth mentioning that we consider here a thermal initial abundance of the RHNs as this can be regarded as a consequence of going through a strong washout regime.  The evolutions of number densities of $N$ and $B-L$ asymmetry can be obtained by solving the following set of coupled Boltzmann's equations~(BEQs) \cite{Buchmuller:2004nz, Davidson:2008bu}:
\begin{gather}\label{beq}
 \frac{dN_{N_{i}}}{dz} = - D_i (N_{N_{i}} - N^{eq}_{N_{i}}), ~~~ \text{with}~~~ i = 1,2\\ 
 \frac{dN_{B - L}}{dz} = -\sum_{i = 1}^2 \epsilon_i D_i  (N_{N_{i}} - N^{eq}_{N_{i}})-\sum_{i = 1}^2 W_{i} N_{B - L},
\end{gather}
with $z = \text{Mass of the lightest RHN}/{\rm Temperature}=M_{N_i}/T$, being $N_i$ as the decaying RHN. $W_i$ denotes the contribution to the total washout\footnote{In general $W_i$ contains three LNV processes, (1) the ID, (2) $2\,\leftrightarrow\,2$ scatterings involving top quarks and gauge bosons (see for instance \cite{Plumacher:1996kc}) which violate lepton number by one unit, and (3) the ones which violate lepton number by 2 units. All of them can influence the initial RHN number density. However, for a scenario of strong washout $(\Gamma_{\text{total},i} > H)$, the final asymmetry is insensitive to the initial RHN abundance, and hence is realized by choosing a thermal initial abundance.  However, for completeness we have calculated the reaction densities (using the prescriptions provided in \cite{Plumacher:1996kc}) for lepton number conserving scatterings, e.g., $ NN \rightarrow \chi_1 \chi_1$ and $ NN \rightarrow hh$. These reaction rates seem to freeze out quite early at $z = 5$ in comparison to the reaction density for the ID. This in turn indicates the ID to be the most dominant washout process to be considered\cite{Buchmuller:2004nz}.}  term due to inverse decay~(ID), $\Delta L \,=\,1$, and $\Delta L \,=\,2$ scatterings. It is clear from the above set of coupled equations that, one can get the RHN abundance from the first equation and the later determines the $B-L$ number density which survives in the interplay of the asymmetry production and it's washout, with respect to temperature. To ensure the participation of $N_1,~ N_2$ in creating the final lepton asymmetry, one has to define a temperature-function ($z$) (for a bit detail one may look into \cite{Konar:2020vuu}),  writing $z= z_i/ \sqrt{x_{1i}}$ with $i = 1,2$. In principle,  $N_{N_i}$'s are the comoving number densities normalised by the photon density at temperature larger than $M_{N_i}$. The washout term $W_i$ present in the above set of BEQs (considering ID only) can be expressed as (taken from \cite{Buchmuller:2004nz}),
\begin{equation*}
W_i = \frac{1}{2} \frac{\Gamma _{\rm ID} (z_i)}{H z}
\end{equation*}

 With the Hubble expansion rate $H \approx \sqrt{\frac{8 \pi^3 g_*}{90}}\frac{M_{N_i}^2}{M_\text{Pl}}\frac{1}{z^2}\approx 1.66 g_*\frac{M_{N_1}^2}{M_{\text{Pl}}}\frac{1}{z^2}$ the decay term $D_i$ in Eq.(\ref{beq}) can be cast into \cite{Buchmuller:2004nz},
\begin{equation}\label{decayterm}
D_i = \frac{\Gamma_{\text{total},i}}{H z} = K_i x_{1i}z \frac{\mathcal{K}_1(z)}{\mathcal{K}_2(z)},
 \end{equation}
with $\Gamma_{\text{total},i}$ as the decay rate of the $ith$ RHN which is given by,
\begin{equation}
\label{decaywidth}
\Gamma_{\text{total},i} = \frac{\left(Y_D Y_D^\dagger + Y_S Y_S^\dagger\right)_{ii}}{4\pi}M_{R_i}. 
\end{equation}

In the Eq.~(\ref{decayterm}) $\mathcal{K}_1$ and $\mathcal{K}_2$ mean the modified Bessel functions of the second kind. The washout factor $K_i$ in Eq.(\ref{decayterm}) quantifies the deviation of the decay rate of the RHNs from the expansion rate ($H$) of the Universe and can be read as,
\begin{align}
 K_i \equiv \frac{\Gamma_{\text{total},i}}{ H (T=M_{N_i})}.
 \end{align}

One can express the baryon-to-photon ratio with the help of lepton asymmetry as, 
\begin{equation}
\eta_B = 0.01 N_{ B-L}^{\rm fin} = 0.01 \epsilon_1 \kappa_1^{\rm fin},
\end{equation}
where $\kappa_1^{\rm fin} = \kappa_1 (z \rightarrow \infty) < 1$ is the final efficiency factor.
The final amount of $B - L$ asymmetry can also be parametrized as $Y_{ B-L} = n_{B-L}/s$, where $s = 2\pi^2g^*T^3/45$ being the entropy density and $g^*$ as the effective number of spin-degrees of freedom in thermal equilibrium ($g^* = 110$). After reprocessing by sphaleron transitions, the baryon asymmetry is related to the $B - L$ asymmetry by $Y_B = (12/37) (Y_{B-L}$)~\cite{Plumacher:1996kc}.

\section{Results and analysis}\label{results}
\subsection{Parameter space for leptogenesis}
\begin{figure}[h!]
\begin{center}
\includegraphics[scale=0.32]{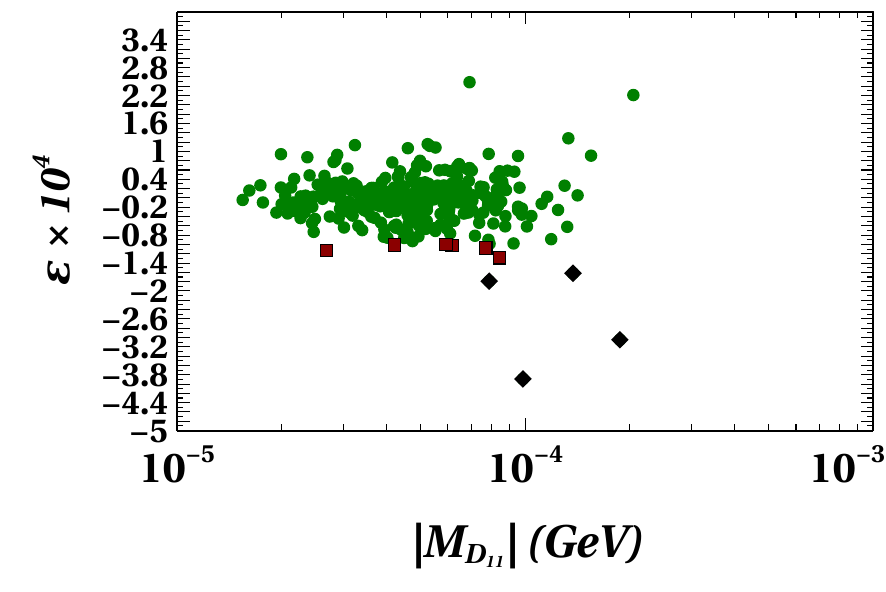}
\includegraphics[scale=0.32]{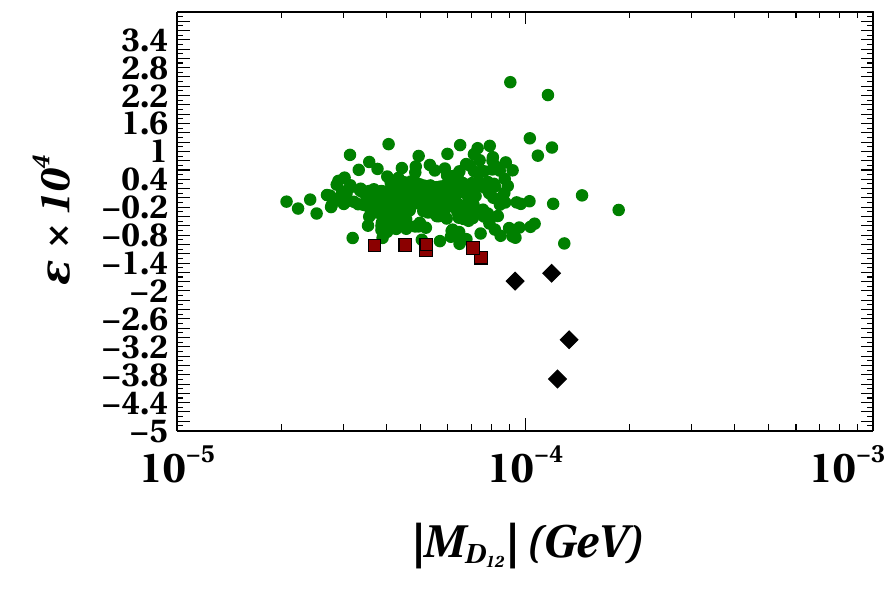} 
\includegraphics[scale=0.32]{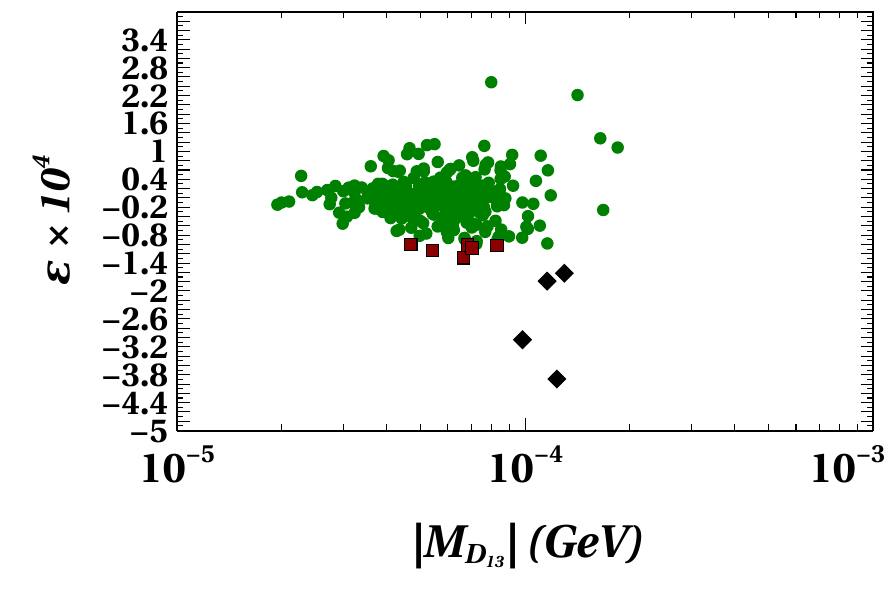}  \\    
\includegraphics[scale=0.32]{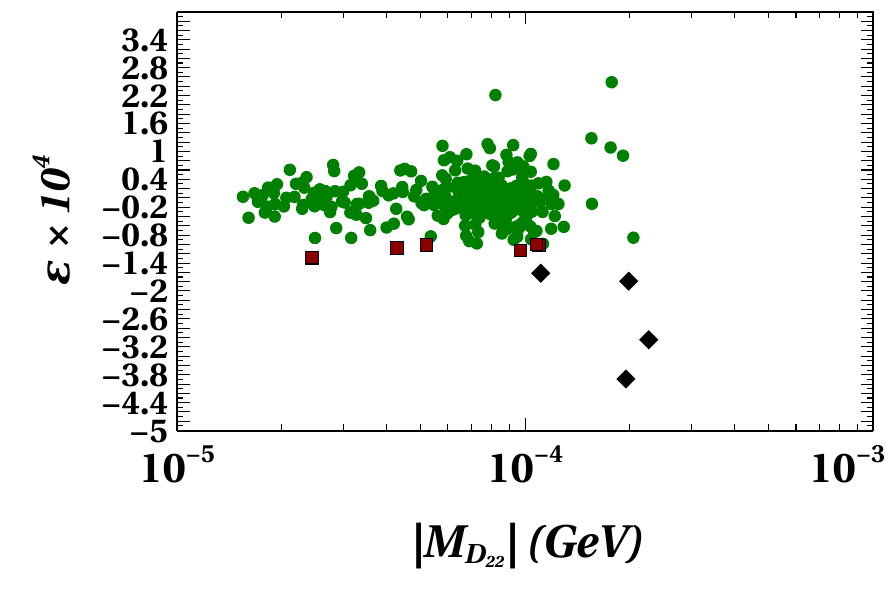} 
\includegraphics[scale=0.32]{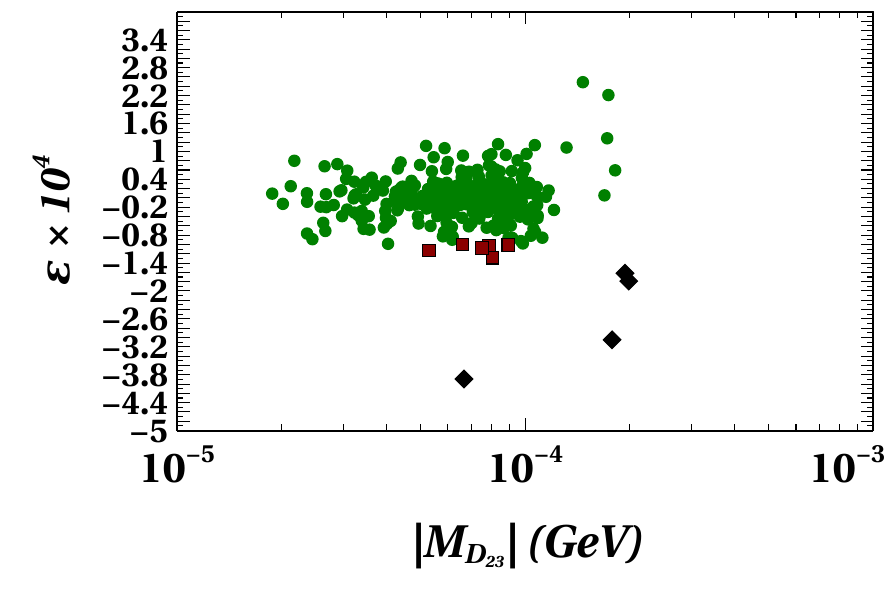} 
\includegraphics[scale=0.32]{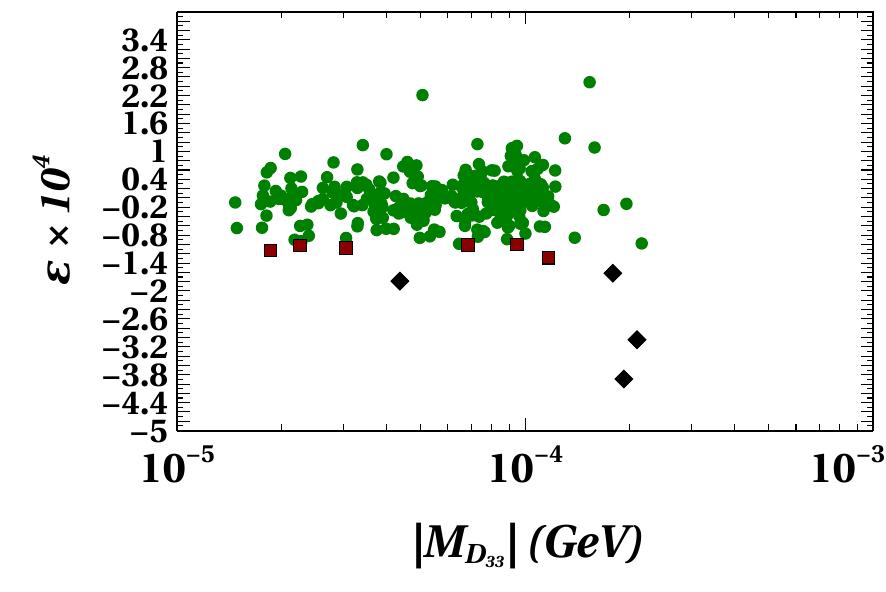}   
\caption{Lepton asymmetry versus matrix elements of $M_D$. The red squares are identified with the appropriate parameter space which is essential for a successful leptogenesis. The black color indicates those parameter spaces which after having an adequate lepton asymmetry can not successfully account for the observed $\eta_B$.}
\label{mdepsilon}
\end{center}
\end{figure}

\begin{figure}[h!]
\begin{center}
\includegraphics[scale=0.32]{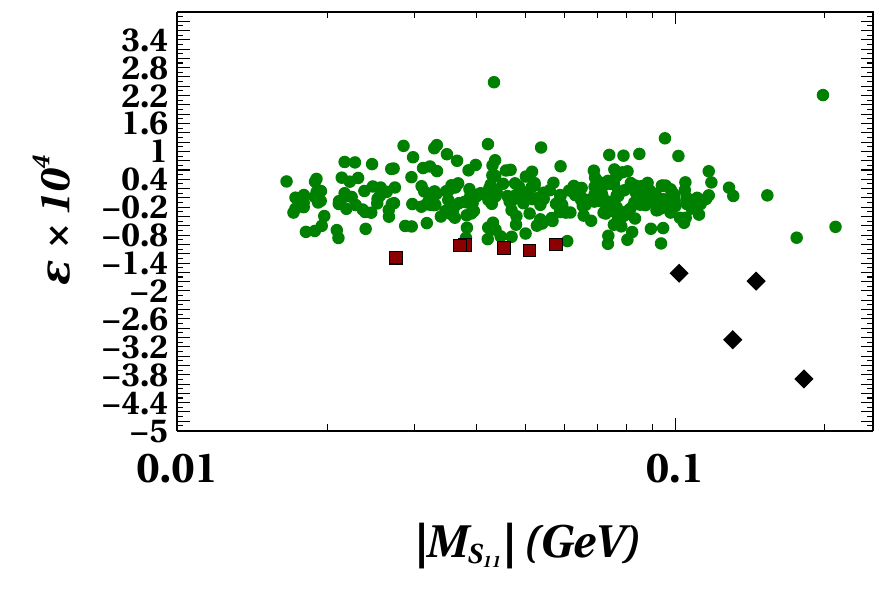}
\includegraphics[scale=0.32]{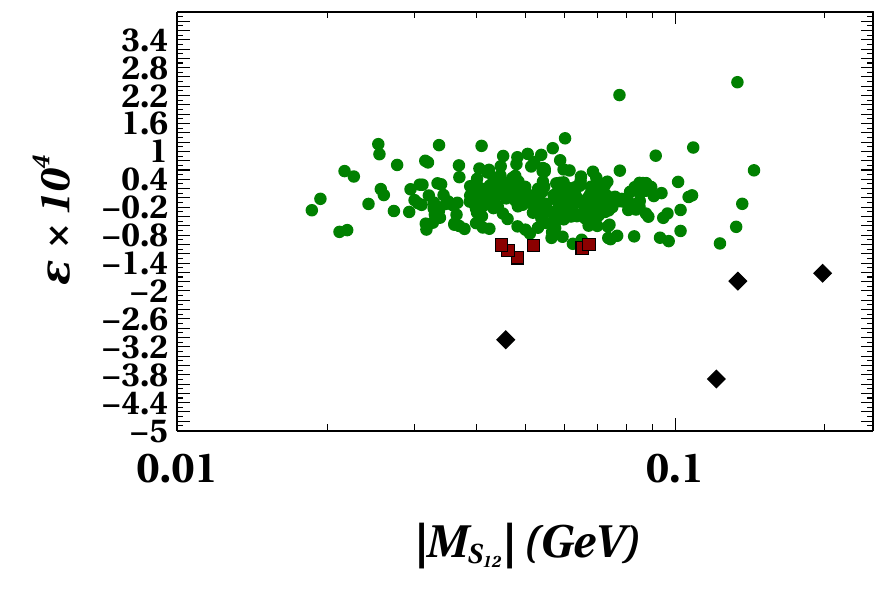} 
\includegraphics[scale=0.32]{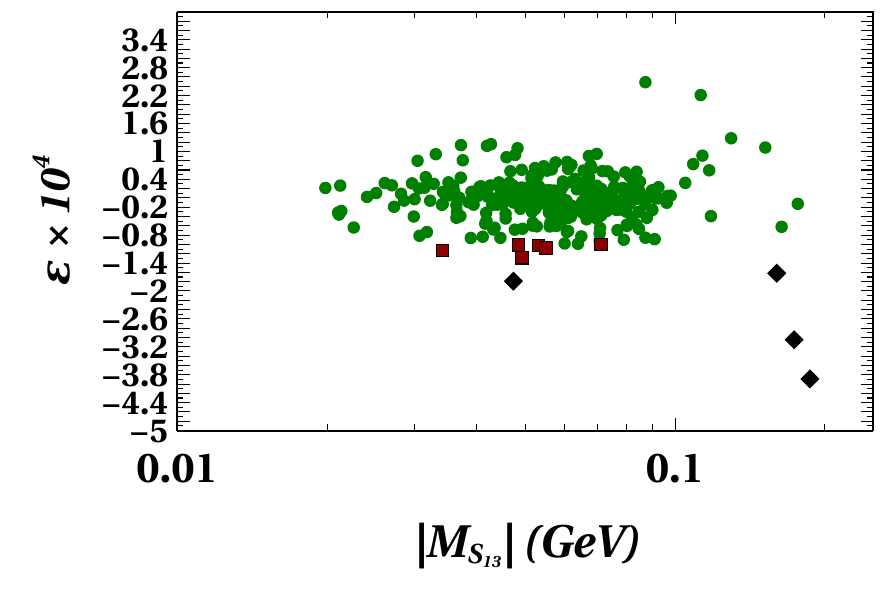}  \\    
\includegraphics[scale=0.32]{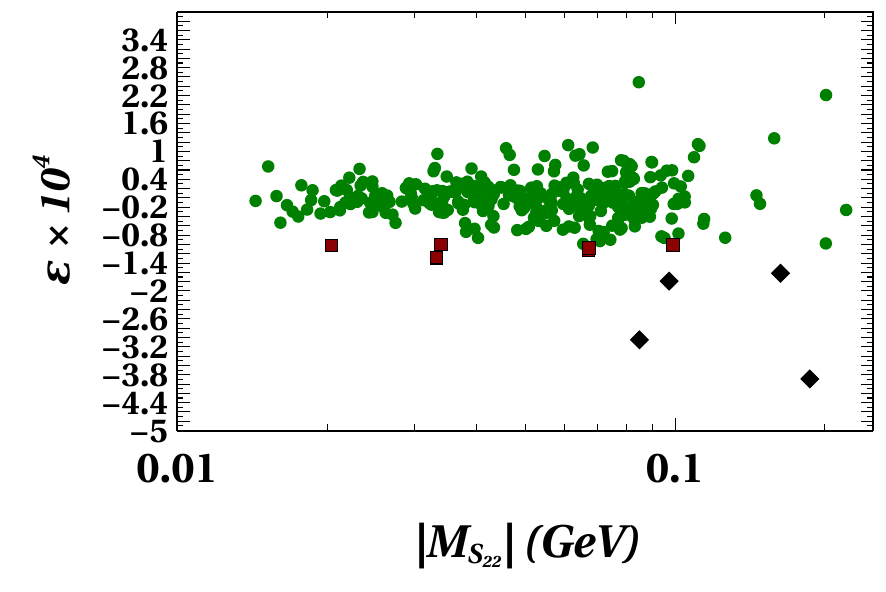} 
\includegraphics[scale=0.32]{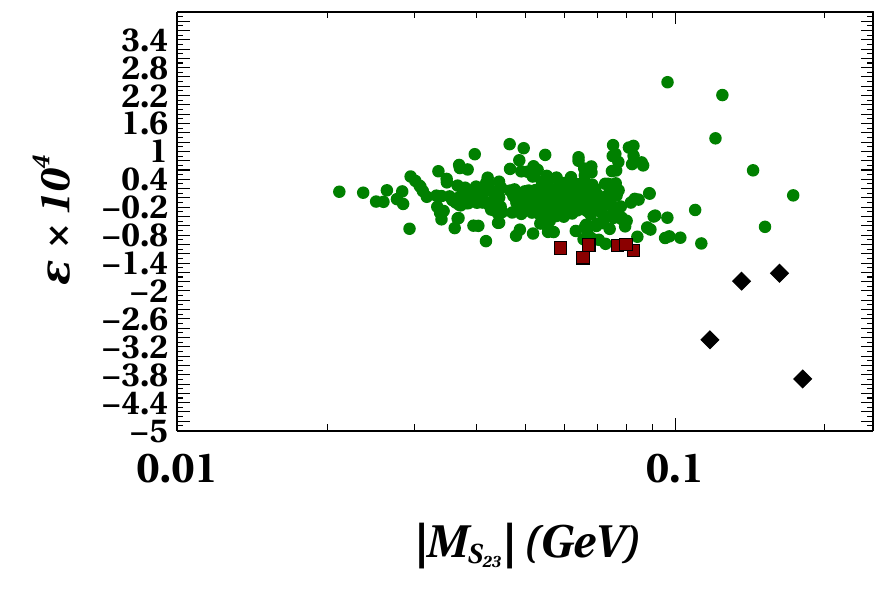} 
\includegraphics[scale=0.32]{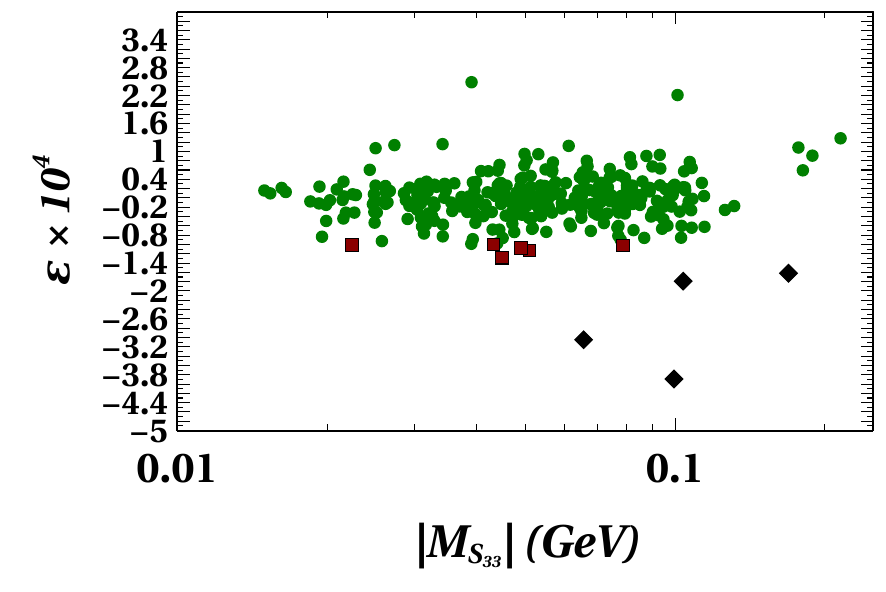}   
\caption{Lepton asymmetry versus matrix elements of $M_S$. To have an idea about the various colored data points one can refer to the caption of fig.~\ref{mdepsilon}.}
\label{msepsilon}
\end{center}
\end{figure}

It is important to provide an order of magnitude estimate of the lepton asymmetry obtained with the help of the model parameters that essentially satisfies all the constraints in the neutrino sector. As evident from the Eq. \eqref{leptonasymmetry}, there exist essentially two different sets of Yukawa couplings namely~($Y_D,~ Y_S$), which determine the order of the lepton asymmetry one should get in this ESS framework. 
These two complex Yukawa couplings are the sources of lepton asymmetry in the present scenario. We show the role of these complex coupling coefficients in the form of the corresponding matrix elements in obtaining the correct order of lepton asymmetry $\epsilon$ in fig.~\ref{mdepsilon}, and fig.~\ref{msepsilon}. These two figures emphasize on the choices of Yukawa couplings and in turn the ESS parameter space in order to make this framework suitable for leptogenesis. For this purpose different color codes have been used to highlight the ESS parameter spaces that are essentially required to be chosen (by dark red square) and omitted (by black vertical-square). 
It is important here to note from the expression of the lepton asymmetry (Eq. \eqref{leptonasymmetry}) that, the large Yukawa coupling $Y_S$ plays a vital role in bringing the  order of lepton asymmetry even up to around $10^{-3}$ for a considerably large region of the entire parameter space of ESS. In addition to this, we would like to emphasize on the fact that, despite of this large $Y_S$ the criteria satisfying $|M_i - M_j| \sim \frac{\Gamma_j}{2}$ is unavoidable even after the presence of an additional self-energy diagram in the present set-up. This is guided by keeping a quasi-degenerate RHN mass spectrum for instance $M_1\approx M_2 <M_3$.  Thereafter,  we consider the decay of more than one RHN and have noticed that the generated asymmetry is solely dependent on the lightest heavy RHN ($N_1$ here), even if the second RHN ($N_2$) is not having pronounced mass hierarchy with the former. For a clear understanding of this fact one may refer to table \ref{tab:etab}. In fig.~\ref{mdepsilon} we show the variation of the lepton asymmetry as a function of the matrix elements $M_D$. The reliance of the lepton asymmetry on the  $M_S$ matrix elements have been shown in fig.~\ref{msepsilon}. This is also evident from the table \ref{tab:coupling}, where we present the Yukawa couplings for the particular benchmark value of the lepton asymmetry parameter. Presented data in the table~\ref{tab:etab}~  leads to the realization that in this ESS scenario one has to be very careful while choosing the texture of Yukawa coupling matrices, as it is evident that even after having a large magnitude of lepton asymmetry one can not reproduce the observed baryon to photon ratio since the associated washouts corresponding to those Yukawa couplings are large enough to erase a considerable part of the generated lepton asymmetry. It will be more transparent if one notices the data provided for BP-IIIa in the table~\ref{tab:etab}. The large washout restricts the parameter spaces for the Yukawa couplings in order to have a successful leptogenesis.
\begin{table}[t]
\begin{center}
\small
\begin{tabular}{| c | c | c |}
\hline
\multirow{2}{*}{BP V} & $Y_{S_{ij}} $&  $ 10^{-6}\left(
\begin{array}{ccc}
 2.08\, +1.79 i & 2.26\, +4.26 i & 4.58\, +1.80 i \\
 2.26\, +4.26 i & 1.36\, +3.02 i & 4.79\, +4.43 i \\
 4.58\, +1.80 i & 4.79\, +4.43 i & 2.07\, +3.98i \\
\end{array}
\right)$ \\
\hline
& $Y_{D_{ij}}$ & $ 10^{-7}\left(
\begin{array}{ccc}
 4.61\, +1.49 i & 3.32\, +2.71 i & 3.4788\, +1.58 i \\
 3.32\, +2.71 i & 1.10\, +0.85 i & 2.02\, +4.16 i \\
 3.47\, +1.58 i & 2.02\, +4.16 i & 5.48\, +3.86 i \\
\end{array}
\right)$\\
\hline 
\multirow{2}{*}{BP IIIa} & $Y_{S_{ij}} $& $10^{-6}\left(
\begin{array}{ccc}
 7.36\, +7.03 i & 3.93\, +19.395 i & 4.75\, +15.28 i \\
 3.93\, +19.39 i & 12.08\, +10.92 i & 10.75\, +12.12 i \\
 4.75\, +15.2811 i & 10.75\, +12.12 i & 15.69\, +6.27 i \\
\end{array}
\right)$ \\
\hline
& $Y_{D_{ij}}$ & $ 10^{-7}\left(
\begin{array}{ccc}
 7.30\, +2.98 i & 3.82\, +5.68 i & 6.51\, +3.60 i \\
 3.82\, +5.68 i & 3.88\, +5.04 i & 3.02\, +10.70 i \\
 6.51\, +3.60 i & 3.02\, +10.70 i & 10.10\, +1.78 i \\
\end{array}
\right)$   \\
\hline
\end{tabular}
\caption{Numerical estimates of the two Yukawa coupling matrices which correspond to the benchmark values (BP: V and IIIa) presented in table \ref{tab:etab}.}
\label{tab:coupling}
\end{center}
\end{table}

\begin{figure*}[h!]
\begin{center}
\includegraphics[scale=0.5]{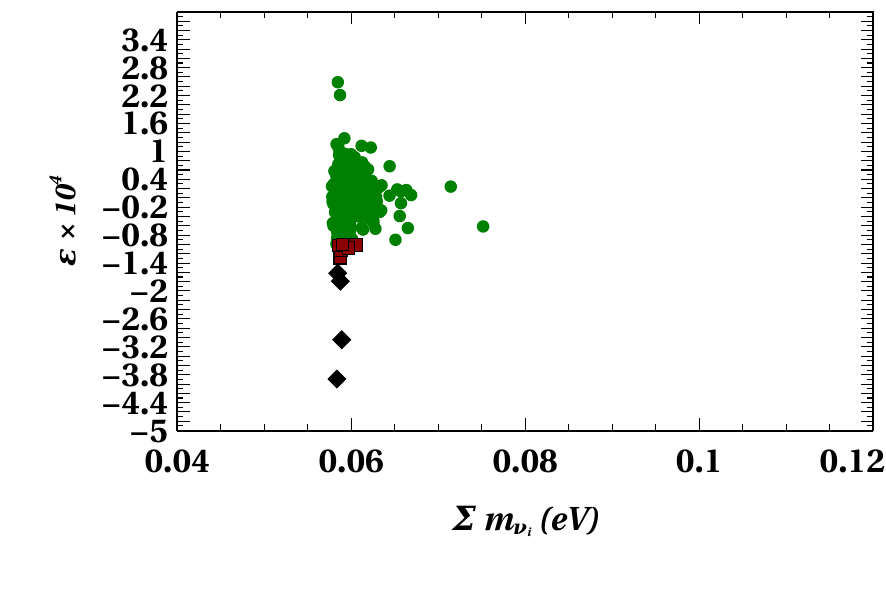}
\includegraphics[scale=0.5]{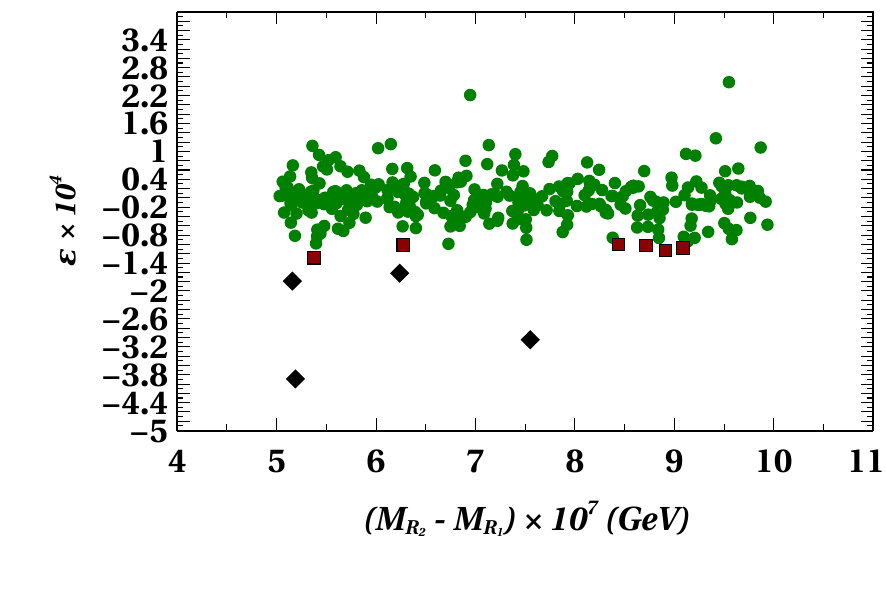}
\caption{Lepton asymmetry as a function of $\sum_{m_{{\nu}_i}}$ and the RHN mass splitting $M_{R_2}-M_{R_1}$. }
\label{epsilon1}
\end{center}
\end{figure*}
\begin{figure*}[h!]
\begin{center}
\includegraphics[scale=0.22]{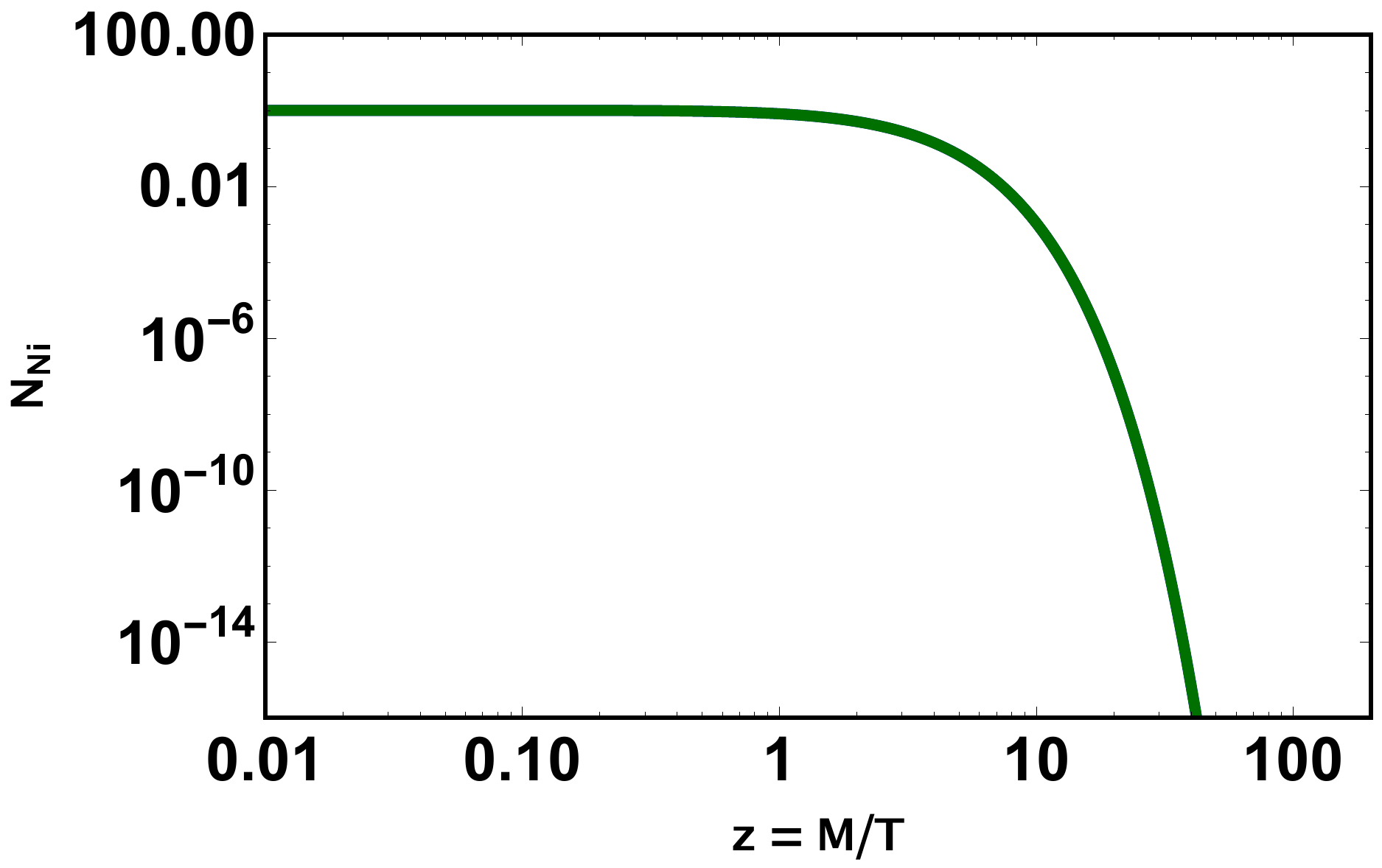}   
\includegraphics[scale=0.22]{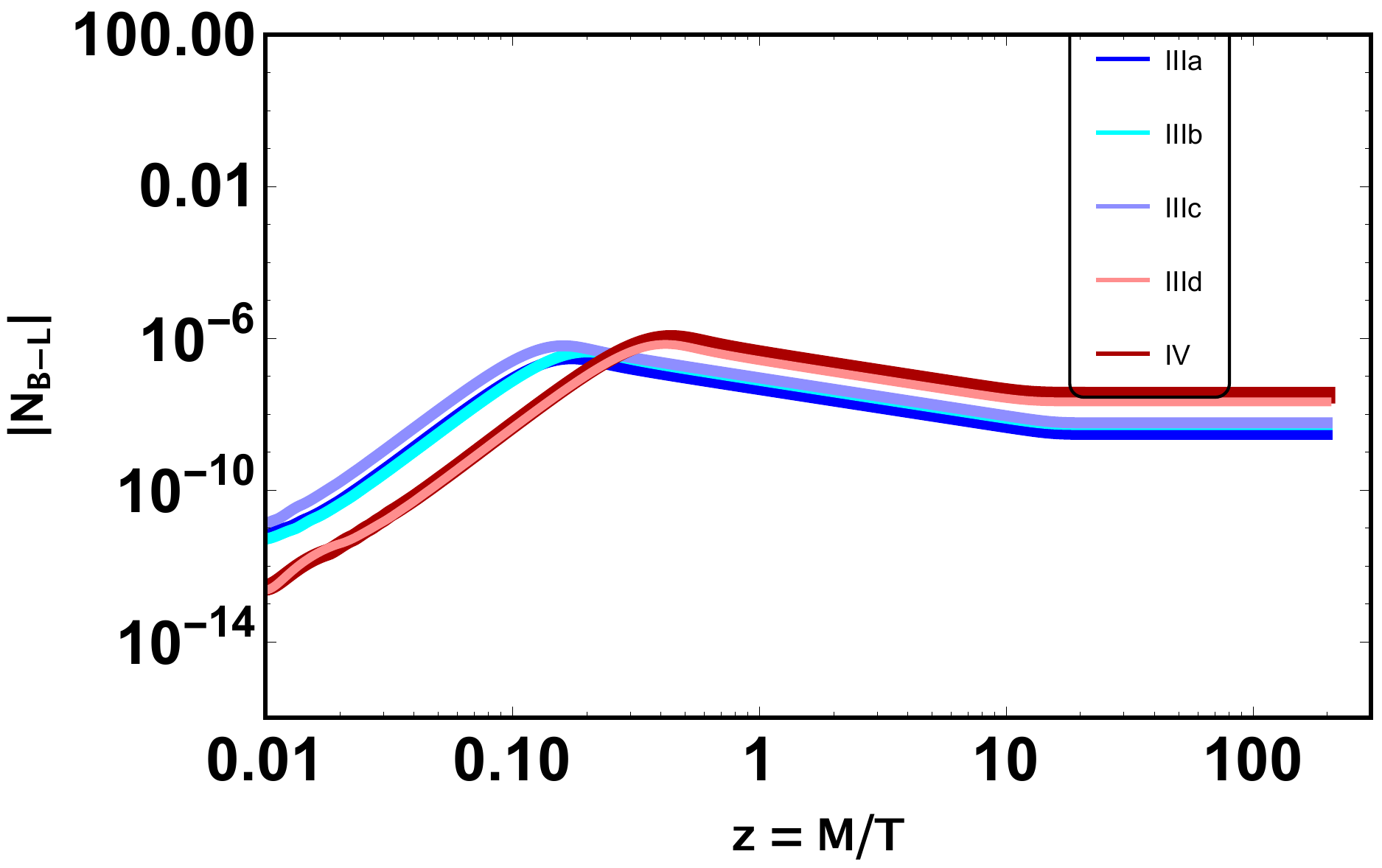}  
\caption{RHN abundance (left panel) and $B-L$ abundance (right panel). Various colors in the second figure represent the final $B-L$ production amount out of each case with different amounts of lepton asymmetries and corresponding washouts,  identified with the benchmark points as tabulated in table~\ref{tab:etab}.}
\label{nbl}
\end{center}
\end{figure*}

In fig.~\ref{epsilon1}~ we present the dependence of lepton asymmetry on the low energy parameters and also on the RHN mass splitting $\Delta M = M_{R_2}-M_{R_1}$.  As it is seen from the left panel of fig.~\ref{epsilon1}~ that the largest asymmetry is obtained for the sum over neutrino mass to fall around $0.06 $ eV. 
\begin{table}[h!]
\begin{center}
\begin{tabular}{| c | c | c | c | c | c | c | c | }
  \hline
BP & $ K_1$ &$K_2$ &$|\epsilon_1|$ & $|\epsilon_2|$ & $\eta_B$   \\
  \hline \hline
 I &  414.62 & 272.60 & $9.52\times 10^{-7}$  & $2.32\times 10^{-15}$   & $2.16\times10^{-12}$ \\
    \hline
    II &  498.9 & 233.85 & $8.3\times 10^{-6}$  & $2.05\times 10^{-15}$   & $1.7\times10^{-11}$ \\
  \hline
  III(a) &  3311.41  & 4052.22 & $1.61 \times 10^{-4}$ & $8.01\times 10^{-15}$     & $2.77\times 10^{-11}$   \\ 
  \hline
  III(b) & 2199.18   & 2452.32  & $  1.78\times 10^{-4}$ & $2.22\times 10^{-15}$     & $5.04\times 10^{-11}$   \\ 
  \hline
  III(c) &  4362.4  & 4343.9 & $3.89 \times 10^{-4}$ & $5.95\times 10^{-15}$     & $5.57\times 10^{-11}$   \\ 
  \hline
    III(d) &  1795.48  & 850.57 & $3.04 \times 10^{-4}$ & $2.85\times 10^{-15}$     & $1.9\times 10^{-10}$\\
  \hline
  IV &  182.15 &  415.42 & $1.13\times 10^{-4}$  & $1.18\times 10^{-15}$   & $3.003\times10^{-10}$ \\
  \hline
    V &  284.47& 395.30 & $1.28\times 10^{-4}$  & $2.05\times 10^{-15}$   & $2.95\times10^{-10}$ \\
    \hline
\end{tabular}
\caption{Bench mark values for the lepton asymmetry parameters and the corresponding washout amount which altogether yield the final baryon to photon ratio ($\eta_B)$. These $\eta_B$ values are evaluated at $z = 100$ from the solution of the Boltzmann Equations.}
\label{tab:etab}
\end{center} 
\end{table}
The amount of the washout we get in this scenario turns out to be of the order of $10^{2, 3}$ which is in agreement with the situation one has to go through while having a resonantly enhanced lepton asymmetry. Moreover, this ensures a strong washout ($K \gg 1$) regime which further leads to the conclusion that the final baryon asymmetry is independent of the initial conditions \cite{Buchmuller:2004nz}. The initial condition here implies an initial abundance of the RHNs which are supposed to decay at a later time. In the weak washout regime ($K<1$) the final baryon asymmetry sensitively depends on the initial number densities of RHNs. In such a picture one has to consider different $\Delta L =1$ scatterings (mentioned in \cite{Plumacher:1996kc}), which may have certain $B-L$ interactions in the context of the particle content present here. The final residual $B-L$ asymmetry caused by the interplay of creation of the asymmetry and its washout through the concerned processes, which in this case is inverse decay, are presented in fig.~\ref{nbl}. One can note here that  inverse decays are sufficient to take over the total washout in a scenario with such huge washout. Moreover, the model parameter space of this ESS model strictly favors a strong washout regime ensuring that one can simply ignore the $2\,\leftrightarrow\,2$ scatterings while numerically solving the set of BEQs setting the initial condition on the RHN abundance as $N^i_{N_1}\,=\,N^{\text{eq}}_{N_1}$. 
%
In the right panel of fig.~\ref{nbl} we present the $B-L$ evolutions with respect to the temperature function for different values of lepton asymmetry which are large in number. One can notice from this figure that, even after having a large amount of lepton asymmetry we may not generate the required amount of $B-L$ charge if the corresponding washouts erase a part of the asymmetry leading to a decrease in the final baryon asymmetry. This fact can also be well-understood from the table~\ref{tab:etab}.

\section{Neutrinoless double beta decay}\label{ndbd}
Neutrinoless double beta decay ($0\nu\beta\beta$) is a lepton number violating process, which, if observed, would establish the Majorana nature of neutrinos with certainty~\cite{Majorana:1937vz, Racah:1937qq, Furry:1939qr, Rodejohann:2011mu}.  In various beyond SM scenarios the rate of this process can get enhancement and become observable in experiments~\cite{Rodejohann:2011mu, Vergados:2016hso, DellOro:2016tmg}. The dominant contribution to the $0\nu\beta\beta$-decay rate, $\Gamma_{0 \nu}$, is due to the exchange of three light Majorana neutrinos $\nu_i$~\cite{Bilenky:2012qi}. However it is customary to look for other contributions to the final process owing to the presence of extra particle species that can mediate the said process. In ESS, additional sterile neutrinos can play important role. 
To this end, along with the light Majorana neutrinos $\nu_i$, the right-handed (RH) neutrino states $N_{R_i} ~\rm{for}~ (i=1,2,3)$ and $S_{L_j} ~\rm{for}~ (j=1,2,3)$ can also mediate $0\nu\beta\beta$ process $2n \rightarrow 2 p + 2 e^{-}$. Due to its heavier mass the contribution of $N_{R_i}$ to the final amplitude of the $0\nu\beta\beta$ process will be suppressed enough, as the strength of this particular channel will be determined by the mixing of active neutrino states with the heavy sterile ($V_{eN_i}$). Additionally, the contribution will be suppressed by the heavy sterile mass $M_{{R}_{i}}$. The $V_{eN}$ is given by ${M^{\dagger}_{D}}{M^{-1}_{R}} W_{N}$ (see Eq.~\ref{eq:Ufinal}) where $W_{N}$ is defined as diagonalization matrix of heavy sterile states given in Eq.~\ref{eq:diangonalizationU2}. The contribution from the $0\nu\beta\beta$ channel mediated by the heavy sterile neutrino can be expressed by 
\begin{equation}
\label{Ncontbtn}
\mathcal{A}_{N} \sim  \frac{{V}^2_{eN_i}}{M_{R_i}}
\end{equation}
where $V_{eN_{i}}$ is the mixing of active neutrinos with the $N_{R_i}$ states.
On the other hand the amplitude of the $0\nu\beta\beta$ process, mediated by the comparatively light sterile neutrino states $S_{L_i}$ for $m_{{s}_{i}} > |\langle p^{2} \rangle|$ can be expressed as 
\begin{equation}\label{Scontbtn}
\mathcal{A}_{S} \sim  \frac{{V}^2_{eS_{i}}}{m_{{s}_{i}}},
\end{equation}
with $V_{eS_{i}}$ being the mixing between the active neutrino with the sterile neutrino given by $M_{D}^{\dagger}{M_{S}^{-1}}^{\dagger}W_{s}$ in the present framework, whereas, 
for  $m_{{s}_{i}} < |\langle p^{2} \rangle|$ 
\begin{equation}
\label{e:amplS}
\mathcal{A}_{S} \sim  \frac{{V}^2_{eS_i}m_{{s}_{i}}}{\langle p^{2} \rangle}
\end{equation}
One can write the simplified expression for the amplitude, taking simultaneously into account $|\langle p^{2} \rangle|  \simeq m_{s_i}^{2} \simeq 100 - 200\, \rm{ MeV}^2$. 
\begin{equation}
\label{e:amplS_small}
\mathcal{A}_{S} \sim  \frac{{V}^2_{eS_i} m_{s_i}}{\langle p^{2} \rangle -m_{{s}_{i}}^{2}}.
\end{equation}
The  half-life of $0\nu \beta \beta$ is given by
\begin{equation} \label{e:Thalf1}
\left(T_{1/2}^{0\nu}\right)^{-1} = K_{0 \nu} \left| \frac{U_{ei}^2  m_{\nu_i}}{\langle p^{2} \rangle- m_{\nu_i}^{2}}  +  \frac{{V}^2_{eS_i} m_{s_i}}{\langle p^{2} \rangle -m_{s_{i}}^{2}}  \right|^{2},
\end{equation}
where, $i$ implies for all the light and heavy neutrino states, supposed to mediate the potential $0\nu \beta \beta$ process. The details of the notation and convention can be found in \nameref{sec:appenA}. In the above, we also have  $K_{0\nu} = G_{0\nu} (\mathcal{M}_{N} m_{p})^{2}$ and  
$\langle p^{2} \rangle\equiv -m_{e} m_{p} \frac{\mathcal{M}_{N}}{\mathcal{M}_{\nu}}$~\cite{Mitra:2011qr}. Here $G_{0\nu}$ is the phase space factor, $\mathcal{M}_{N} $ is the nuclear matrix element (NME) for heavy sterile exchange whereas $\mathcal{M}_{\nu}$ is the corresponding element for light sterile exchange. The required values of $G_{0\nu}$ and NMEs have been taken from Ref.~\cite{Meroni:2012qf}. In our analysis, we stick to particular values of NMEs as $\mathcal{M}_{N}$ = 163.5 and $\mathcal{M}_{\nu}$ = 2.29; and the phase space factor $G_{0\nu}$ is $5.92 \times 10^{-14}~ {\rm year}^{-1}$.
\begin{figure*}[h!]
\begin{center}
\includegraphics[scale=0.5]{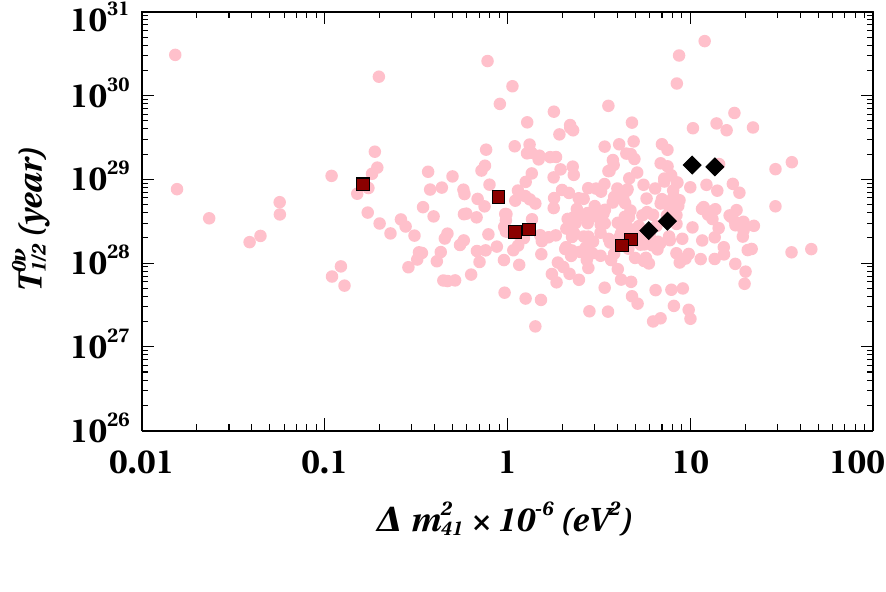}
 \includegraphics[scale=0.5]{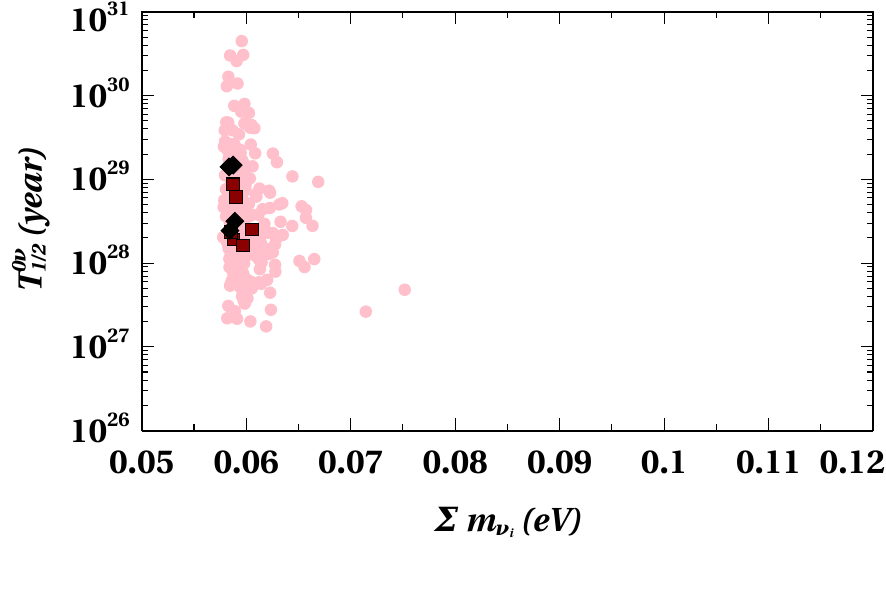} 
\caption{Prediction for the half-life of the $0 \nu \beta \beta$ process as a function of the fourth mass squared splittings $\Delta m^2_{41}$ (left) and sum over neutrino masses (right). The reason behind the choice of different colors can be obtained from the caption of fig.~\ref{sterilemixing}. Here the values of nuclear matrix elements (NMEs) are $\mathcal{M}_{N}$ = 163.5 and $\mathcal{M}_{\nu}$ = 2.29; the phase space factor  $G_{0\nu}$ is taken as $5.92 \times 10^{-14}~ {\rm year}^{-1}$~\cite{Meroni:2012qf}.}
\label{fig:ndbd}
\end{center}
\end{figure*}
For completeness we focus on the contribution of the lighter sterile neutrino states $S_{L}$. As mentioned earlier, the heavier states ($N_R$), being considerably heavier $~10^{4}-10^{5}$ GeV and having the active-sterile mixing $\sim M_{D}/M_{R} \sim 10^{-10}- 10^{-8}$ give negligible contribution in $0\nu \beta\beta$.  On the other side there seems to be sizeable contribution from the the lighter sterile neutrino states to the final amplitude of $0\nu \beta \beta$ process, due to the larger active-sterile mixing $V_{eS}$.  It is worth mentioning here that taking care of all the constraints discussed above, the mass range of that lighter sterile states ($S_L$) lies in the keV range.

The theoretical prediction on the $T_{1/2}^{0\nu}$ drawn from the constraints of this framework, has been compared with the one reported by the recent  KamLAND-Zen experiment which sets $T_{1/2}^{0\nu} > 1.07 \times 10^{26}$ year~\cite{KamLAND-Zen:2016pfg, Penedo:2018kpc}.  Till date the best lower limit on the half-life of the $0\nu \beta\beta$ using $^{76}$Ge is $ T^{0\nu}_{1/2} >8.0\times 10^{25}$ yrs at 90$\%$ C.L. from GERDA \cite{Agostini:2018tnm}.  
Though we do not have any direct experimental proof for $0\nu \beta \beta$ transition~\cite{KlapdorKleingrothaus:2000sn, Aalseth:2002rf, Argyriades:2008pr, Arnaboldi:2002du, KlapdorKleingrothaus:2006ff, Arnaboldi:2008ds, Sarazin:2000xv, Abt:2004yk, KlapdorKleingrothaus:2004wj} the search for improving the bound on the $T_{1/2}^{0\nu}$ value is still beneficial to explore the possibility of new physics~\cite{Abt:2004yk,Arnaboldi:2002du,Conti:2003av,Guiseppe:2008aa,Arnold:2010tu,Beeman:2011yc,Bloxham:2007aa,Zuber:2001vm,Granena:2009it}.  
The theoretical prediction of the $T_{1/2}^{0\nu}$ value in the ESS framework is shown in  fig.~\ref{fig:ndbd}. It is expressed as functions of square of mass splitting of $S_{1}$ and active electron neutrino and also as function of sum over active neutrino masses $\Sigma m_{{\nu}_{i}}$. It is evident from the figure that all points are allowed from the limit set by KamLAND-Zen experiment~\cite{KamLAND-Zen:2016pfg}.

For completeness, we comment on the effective Majorana mass ($m_{\rm eff}$) in the ESS framework.
The Eq.~\ref{e:Thalf1} can be expressed as~\cite{Nayak:2013dza,Goswami:2020loc}
\begin{eqnarray}\label{e:meffV1}
\left(T_{1/2}^{0\nu}\right)^{-1} 
 &=& \frac{K_{0 \nu}}{|\langle p^{2} \rangle|^{2}} \left| U_{ei}^2  m_{\nu_i}  +  {V}^2_{eS_i} m_{s_i}  \right|^{2},~~{\rm if}~~ \langle p^{2} \rangle \gg m_{s_i}^{2}, \nonumber \\
&=& \frac{K_{0 \nu}}{|\langle p^{2} \rangle|^{2}} \left| U_{ei}^2  m_{\nu_i}  -  {V}^2_{eS_i} \frac{|\langle p^{2} \rangle|}{m_{s_i}}  \right|^{2},~~{\rm if}~~ m_{s_i}^{2} \gg \langle p^{2} \rangle,\nonumber \\
\Rightarrow \left(T_{1/2}^{0\nu}\right)^{-1} &=& G_{0\nu}\left(\frac{\mathcal{M}_{\nu}}{m_{e}}\right)^{2} \left|(m_{\nu}^{ee} + m_{s}^{ee}) \right|^{2} \equiv
G_{0\nu}\left(\frac{\mathcal{M}_{\nu}}{m_{e}}\right)^{2} \left|m_{\rm eff}\right|^{2},
\end{eqnarray}
where $m_{\rm eff} = m_{\nu}^{ee} + m_{s}^{ee}$, and
\begin{eqnarray}\label{e:meffV2}
m_{\nu}^{ee} &=& \sum_{i} U_{ei}^2  m_{\nu_i}, \nonumber \\
m_{s}^{ee} &=& \sum_{i} {V}^2_{eS_i} m_{s_i},~{\rm for}~\langle p^{2} \rangle \gg m_{s_i}^{2}, \nonumber \\
&=& \sum_{i}{V}^2_{eS_i} \frac{|\langle p^{2} \rangle|}{m_{s_i}},~{\rm for}~m_{s_i}^{2} \gg \langle p^{2} \rangle.\nonumber
\end{eqnarray}
From the Eq.~\ref{e:meffV1}, $m_{\rm eff}$ can also be expressed in terms of $T_{1/2}^{0\nu}$ as,
\begin{equation} \label{e:meffV4}
m_{\rm eff} = \frac{1}{\sqrt{T_{1/2}^{0\nu}G_{0\nu}}}\left(\frac{m_{e}}{\mathcal{M}_{\nu}}\right).
\end{equation}
So, for the highest NME value we get the lowest value of $m_{\rm eff}$ and vice versa; $m_{\rm eff}$ decreases with an increase in $T_{1/2}^{0\nu}$. Evidently, the highest value of the NME for the light sterile exchange $\mathcal{M}_{\nu}$ yields the tightest constraint on $m_{\rm eff}$ value.
\begin{figure}[h!]
\begin{center}
\includegraphics[scale=0.5]{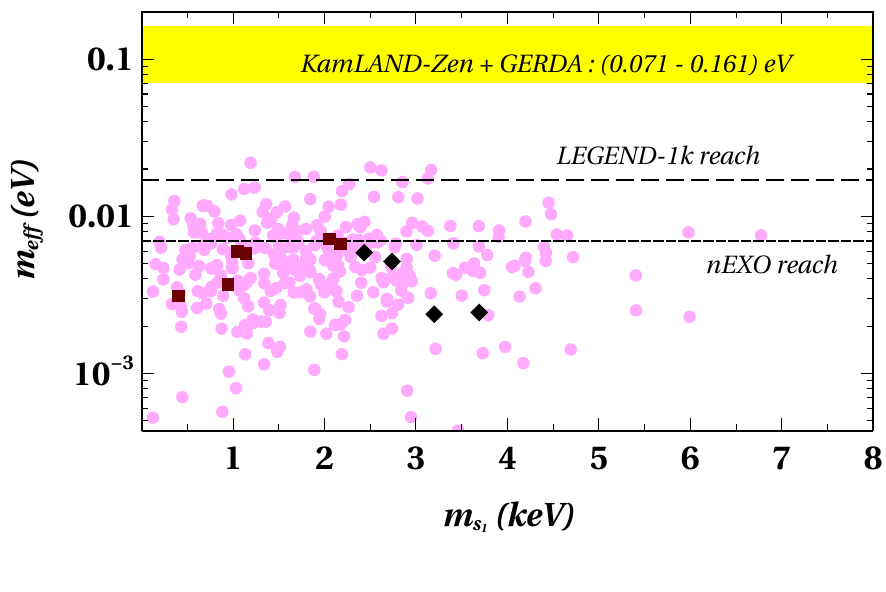}
\includegraphics[scale=0.5]{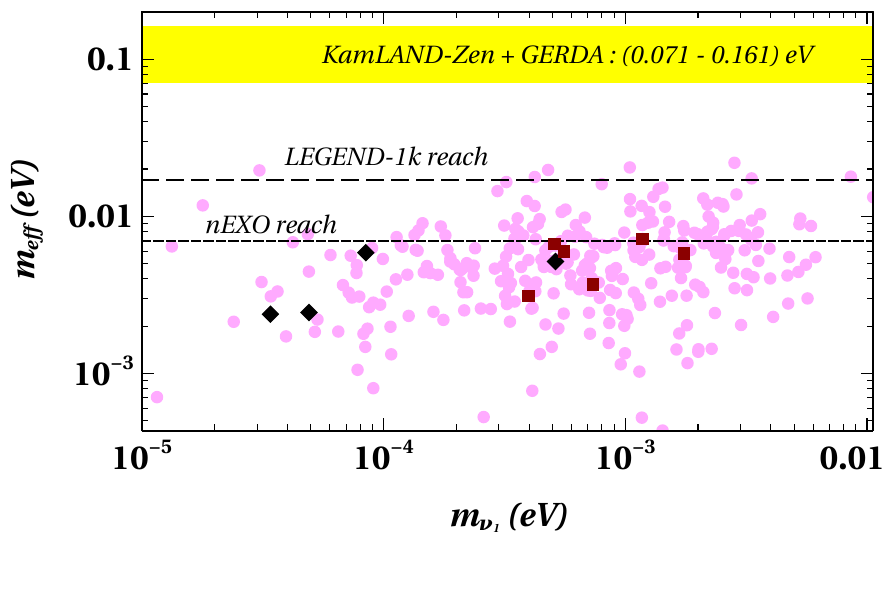} 
\caption{Variation of $m_{\rm eff}$ (in eV) as a function of $m_{s_1}$ (keV) (left panel) and as a function of $m_{\nu_1}$ (eV) (right panel). The choice of different colors can be understood from the caption of fig.~\ref{sterilemixing}. Here the model-value of $m_{\rm eff}$ (eV) has been calculated using $\mathcal{M}_{\nu}$ = 2.29,  $G_{0\nu}$ = $5.92 \times 10^{-14}~ {\rm year}^{-1}$~\cite{Meroni:2012qf}. The region above the yellow band is disallowed from the integrated results of KamLAND-Zen and GERDA~\cite{Agostini:2018tnm, Goswami:2020loc}. The horizontal large-dashed black line represents the future sensitivity of LEGEND-1 $k$~\cite{Agostini:2017jim,LEGEND:2017cdu} whereas the lower one, the small-dashed black horizontal line corresponds to the future sensitivity of nEXO~\cite{nEXO:2018ylp}.}
\label{fig:meff}
\end{center}
\end{figure}
The scattered points in the left and right panel plots of Fig.~\ref{fig:meff} show the calculated value of $m_{\rm eff}$ (eV) of the model as the function of $m_{s_1}$ (keV) and $m_{\nu_1}$ (eV), respectively. The yellow band corresponds to the combined constraints coming from KamLAND-Zen and GERDA~\cite{Agostini:2018tnm, Goswami:2020loc}. The region above this band is disfavored by the aforementioned experiments. It is worth-mentioning that a particular experimental lower bound on the $T_{1/2}^{0\nu}$ yields a band in $m_{\rm eff}$ due to the uncertainty in the NME~\cite{Engel:2016xgb,Kotila:2012zza}.
The proposed 3$\sigma$ future sensitivities of nEXO and LEGEND-1$k$ are $T_{1/2}^{0\nu} = 5.7 \times 10^{27},~4.5 \times 10^{27}$, respectively~\cite{Goswami:2020loc,nEXO:2018ylp,LEGEND:2017cdu}. Considering the highest values of respective NME, we get the lowest values of $m_{\rm eff}$ from nEXO and LEGEND-1$k$ as 0.007 eV and 0.017 eV, which in turn results in the maximum constraint on the model-parameter space~\cite{Goswami:2020loc}. Among the two black horizontal lines, the upper large-dashed one corresponds to the calculated value of $m_{\rm eff}$ from LEGEND-1$k$, where the lower small-dashed one represents the calculated $m_{\rm eff}$ from nEXO. This figure shows that though the estimated $m_{\rm eff}$ from the current experiments does not rule out our parameter space, the proposed sensitivity of future experiments will be able to probe some part of our leptogenesis parameter space.
\section{Conclusion}
\label{conclusion} 
We present a standard model extension augmented by a $U(1)_{B-L}$ symmetry to offer an explanation to the light Majorana neutrino mass as well as the observed matter-antimatter asymmetry of the Universe through a low scale seesaw model. Our study is stimulated by the requirement of having a TeV scale RHN along with a potential dark matter candidate which is of astrophysical significance.  In this work we have systematically investigated the viable parameter space of an extended seesaw framework to meet the observed matter-antimatter asymmetry. Having the ESS framework leads to the realization that the presence of the additional sterile neutrino generations takes a strong hold in the lepton asymmetry parameter space through an extra self-energy contribution. We have emphasized here the new findings which are the prime requirements in order to account for a successful resonant leptogenesis in this ESS framework in the context of the work done by the authors in Ref.~\cite{Kang:2006sn}. We present the key findings of our analysis below. 
The model parameter space arising from this ESS scheme can, in principle, explain both the hierarchies of neutrino mass pattern, however, owing to slight experimental bias towards normal hierarchy, we carry out our analysis in this hierarchy. We have presented mild correlations among the mass matrix elements, which further can explain the properties of the leptonic mixing matrix. In particular it is worth-mentioning that, the allowed parameter space for successful leptogenesis in this ESS model brings out rich predictions on the low energy neutrino mixing data. In the context of leptogenesis we obtained significant preferences in the neutrino observables for the octant of the atmospheric mixing angle ($ \theta_{23}> 45^\circ$) along with the reactor mixing angle ($\theta_{13} >  8.52^\circ$). These low energy predictions for mixing angles make this framework a viable scenario to be tested from the perspective of future and ongoing neutrino oscillation experiments.  This model predicts the entire $3\sigma$ range for the Dirac CP phase $\delta_{\rm CP}$.  It is to be noted here that the allowed values of the Majorana CP phases are found to be in the range $(0^\circ-360^\circ)$ in our analysis. The sum over neutrino mass obtained from this analysis ($\sum_{m_i} < 0.12$~eV) is found to be well within the bound reported by Cosmology \cite{Vagnozzi:2017ovm,Giusarma:2016phn}. We also briefly address on how the model parameter space impacts the half-life period of the $0 \nu \beta \beta$ decay process ($T_{1/2}^{0\nu} >10^{26}$ yrs) which is in agreement with the half-life period as envisaged by the recent KamLAND-Zen experiment. We particularly highlight the results of this analysis in the leptogenesis context as we have obtained drastically different result from the one presented in \cite{Kang:2006sn} where an assumption on some of the Yukawa elements appears to be the essential source of successful leptogenesis. Without resorting to any such assumption, we, on the other hand, calculated them from a thorough scan of the ESS parameter space which satisfy 3$\sigma$ global-fit neutrino oscillation data.

In the later part, we have discussed how the resulting complex Yukawa coupling matrices play their role in realizing baryogenesis through leptogenesis scenario by the CP-violating decay of TeV scale RHN. We keep the RHN mass window to be around 10-20 TeV. It is worth mentioning that in spite of having an additional self-energy contribution to the lepton asymmetry, the tiny splitting among the RHN masses is unavoidable which is a basic criteria of facing resonant leptogenesis. The washout in this framework strictly hints towards a strong washout regime, which is also approved by the low energy neutrino data. The highest and ample amount of lepton asymmetry we report here to be around $1.13 \times 10^{-4}$. We also observe here that although for some parameter space of ESS we obtain a large amount of lepton asymmetry, the final baryon asymmetry gets depleted by a factor due to the large washout.
The present analysis confers a detailed construction of the extended seesaw parameter space.  It is interesting to note that  within the allowed seesaw parameter space satisfying the neutrino oscillation data, only some particular regions can actually account for a sufficient amount of lepton asymmetry which has to be of around $10^{-4}$. Thus it is worth mentioning that this particular extension of the canonical seesaw model play a crucial role in obtaining an enhanced lepton asymmetry which translates into the requirement for the cosmological baryon asymmetry of the Universe both from the quantitative and the qualitative point of view. This model can potentially manifest its signatures in search of TeV scale Majorana right handed neutrino in the future collider  searches. An added bonus of this model is the existence of a keV sterile neutrino which can also play the role of a non-thermal dark matter candidate, the detailed phenomenological study of which we keep for a forthcoming work.

\section*{Acknowledgement} 
UKD acknowledges the support from Department of Science and Technology (DST), Government of India under the grant reference no. SRG/2020/000283. TJ would like to acknowledge the support from Science and Engineering Research Board (SERB), Government of India under the grant reference no. PDF/2020/001053. TJ thanks Ram Lal Awasthi, Sarif Khan, Amina Khatun and Soumya C for important discussions. TJ also acknowledges Adrish Maity, Saptarshi Mukherjee and Tapolina Jha for important discussion regarding the computational part. Authors also acknowledge them as a part of the simulation has been performed using their machine at initial stage of the work. Authors acknowledge the HPC facility (Vikram-100 HPC) provided by PRL, Ahmedabad. AM also wants to thank Ashimananda Modak for useful discussion related to numerical simulation. AM would like to acknowledge the financial support provided by SERB-DST, Govt. of India through the project EMR/2017/001434.  NS acknowledges RUSA 2.0 Project.

\section*{Appendix}
\label{sec:appenA}
Here we briefly explain the diagonalization procedure in extended seesaw framework for completeness. This is largely based on the method outlined in~\cite{Grimus:2000vj, Mitra:2011qr}. The neutral mass matrix $\mathcal{M}_{n}$ is given by
\begin{equation}
\label{eq:extcsawMat}
\mathcal{M}_{n} = \begin{pmatrix} 0 & 0 & M_{D}^{T} \cr 0 & \mu  & M_{S}^{T} \cr M_{D} & M_{S} & M_{R} \end{pmatrix}.
\end{equation}
In the entire analysis, we consider Majorana matrix $M_{R}$ to be real and diagonal. The $\mu$ being another Majorana mass matrix is considered to be complex symmetric. The Dirac mass matrices $M_{S}, M_{D}$ are also complex symmetric matrices. The matrix $\mathcal{M}_{n}$ can be block-diagonalized as ~$\mathcal{U}_{1}^{T}\mathcal{M}_{n}\mathcal{U}_{1} = \mathcal{M}_{bd}$ where, $\mathcal{U}_{1}$ and $\mathcal{M}_{bd}$ are the final block-diagonalization matrix and final block-diagonal mass matrix respectively. Furthermore, $\mathcal{U}_{1}$ can be decomposed as $\mathcal{U}_{1} = \mathcal{U}_{1}^{\prime} \mathcal{U}_{1}^{\prime \prime}$. So, we have ${\mathcal{U}_{1}^{\prime}}^{T}\mathcal{M}_{n}\mathcal{U}_{1}^{\prime} = \mathcal{\hat{M}}_{bd}$ followed by another block-diagonalization as ${\mathcal{U}_{1}^{\prime \prime}}^{T}\mathcal{\hat{M}}_{bd}\mathcal{U}_{1}^{\prime \prime} = \mathcal{M}_{bd}$, with $\mathcal{\hat{M}}_{bd}$ being the intermediate block-diagonalization matrix. We follow the parametrization of Ref. \cite{Grimus:2000vj}, \textit{i.e.},
\begin{equation}
\label{eq:U1prime}
\mathcal{U}_{1}^{\prime} = \begin{pmatrix} \sqrt{\mathbb{1}_{2 \times 2} - \mathcal{B}\mathcal{B}^{\dagger}} &  \mathcal{B}_{2 \times 1} \cr - {\mathcal{B}^{\dagger}}_{1 \times 2} &   \sqrt{\mathbb{1}_{1 \times 1} - \mathcal{B}^{\dagger}\mathcal{B}} \end{pmatrix},
\end{equation}
where, $\mathcal{B} = \begin{pmatrix} \mathcal{B}_{a a} \cr \mathcal{B}_{b b} \end{pmatrix}$, $\mathcal{B} = \mathcal{B}_{1} + \mathcal{B}_{2} + \mathcal{B}_{3} + \ldots $ 
and $\sqrt{1 - \mathcal{B}\mathcal{B}^{\dagger}} = 1 - \frac{1}{2}\mathcal{B}_{1}\mathcal{B}_{1}^{\dagger} - \frac{1}{2} (\mathcal{B}_{1}\mathcal{B}_{2}^{\dagger} + \mathcal{B}_{2}\mathcal{B}_{1}^{\dagger}) - \frac{1}{2} (\mathcal{B}_{1}\mathcal{B}_{3}^{\dagger} + \mathcal{B}_{2}\mathcal{B}_{2}^{\dagger} + \mathcal{B}_{3}\mathcal{B}_{1}^{\dagger} + \frac{1}{4}\mathcal{B}_{1}\mathcal{B}_{1}^{\dagger}\mathcal{B}_{1}\mathcal{B}_{1}^{\dagger}) + \ldots$; so, $\mathcal{B}_{a a\backslash b b} = \mathcal{B}_{a a1\backslash b b1} + \mathcal{B}_{a a2\backslash b b2} + \mathcal{B}_{a a3\backslash b b3} + \cdots $.
\noindent
From the following diagonalization (block)
%
\begin{align}
\label{eq:diagone}
\mathcal{U}_{1}^{\prime T}\mathcal{M}_{n}\mathcal{U}_{1}^{\prime} = \mathcal{\hat{M}}_{bd} =
\begin{pmatrix} 
\mathcal{M}_{\text{light}_{2\times 2}} & \mathbb{0}_{2 \times 1} \\ 
\mathbb{0}_{1 \times 2} & \mathcal{M}_{\text{heavy}_{1\times 1}} 
\end{pmatrix},
\end{align}
we get,
\begin{equation}
\label{eq:cond1}
\mathcal{B}^{T}\widetilde{M}_{L}\sqrt{1 - \mathcal{B}\mathcal{B}^{\dagger}} - \mathcal{B}^{T}\widetilde{M}_{D}^{T}\mathcal{B}^{\dagger} + \sqrt{1 - \mathcal{B}^{T}\mathcal{B}^{\ast}}\widetilde{M}_{D}\sqrt{1 - \mathcal{B}\mathcal{B}^{\dagger}} - \sqrt{1 - \mathcal{B}^{T}\mathcal{B}^{*}} M_{R} \mathcal{B}^{\dagger} = 0,
\end{equation}
where, $\widetilde{M}_{L} = \begin{pmatrix} 0 & 0 \cr 0 & \mu \end{pmatrix}$ and $\widetilde{M}_{D} = \begin{pmatrix} M_{D} & M_{S} \end{pmatrix}$.\\
Considering $\mathcal{B}_{j} \propto \frac{1}{M_{R}^{j}}$  and equating different coefficients of $M_{R}^{j}$ for different values of $j$ from the above equation:
\begin{align}
\label{eq:B1int}
M_{R}\mathcal{B}_{1}^{\dagger} &= \widetilde{M}_{D},\\
\label{eq:B2int}
M_{R}\mathcal{B}_{2}^{\dagger} &= {M_{R}^{-1}}^{\ast}\widetilde{M}_{D}^{\ast}\widetilde{M}_{L}.
\end{align}
Solving above equations we get,
\begin{align}
\label{eq:Bvalues_int}
\mathcal{B}_{aa1} = M_{D}^{\dagger}M_{R}^{-1},~ \mathcal{B}_{bb1} = M_{S}^{\dagger}M_{R}^{-1},~ \mathcal{B}_{aa2} = 0,~ \mathcal{B}_{bb2} = \mu^{\ast}M_{S}^{T}M_{R}^{-2}.
\end{align}
So, the block-diagonalized mass matrices
\begin{footnotesize}
\begin{flalign}
\label{eq:mlight_int}
&\mathcal{M}_{\text{light}_{2\times 2}} = \widetilde{M}_{L} - \frac{1}{2}(\mathcal{B}_{1}^{\ast}\mathcal{B}_{1}^{T}\widetilde{M}_{L} + \widetilde{M}_{L}\mathcal{B}_{1}\mathcal{B}_{1}^{\dagger}) - \widetilde{M}_{D}^{T}(\mathcal{B}_{1}^{\dagger} + \mathcal{B}_{2}^{\dagger}) + \mathcal{B}_{1}^{\ast}M_{R}\mathcal{B}_{2}^{\dagger} & \nonumber \\
&= \begin{pmatrix} -M_{D}^{T}M_{R}^{-1}M_{D} & -M_{D}^{T}M_{R}^{-1}M_{S} -\frac{1}{2}M_{D}^{T}M_{R}^{-2}M_{S}^{\ast}\mu \cr
                  -M_{S}^{T}M_{R}^{-1}M_{D} -\frac{1}{2}\mu M_{S}^{\dagger}M_{R}^{-2}M_{D} & \mu - M_{S}^{T}M_{R}^{-1}M_{S} - \frac{1}{2}(M_{S}^{T}M_{R}^{-2}M_{S}^{\ast}\mu + \mu M_{S}^{\dagger}M_{R}^{-2}M_{D}) \end{pmatrix}.&
\end{flalign}
\end{footnotesize}
and,
\begin{flalign}
\label{eq:mheavy_int}
&\mathcal{M}_{\text{heavy}_{1\times 1}} = M_{R} + \mathcal{B}_{1}^{T}\widetilde{M}_{L}\mathcal{B}_{1} + \frac{1}{2}\widetilde{M}_{D}\mathcal{B}_{1} + \frac{1}{2}\widetilde{M}_{D}\mathcal{B}_{2} + \mathcal{B}_{1}^{T}\widetilde{M}_{D}^{T} + \mathcal{B}_{2}^{T}\widetilde{M}_{D}^{T}& \nonumber \\ 
&~~~~~~~~~~~~~~~~ -\frac{1}{2}M_{R}\mathcal{B}_{2}^{\dagger}\mathcal{B}_{1} - \frac{1}{2}\mathcal{B}_{1}^{T}\mathcal{B}_{1}^{\ast}M_{R} - \frac{1}{2}\mathcal{B}_{1}^{T}\mathcal{B}_{2}^{\ast}M_{R} - \frac{1}{2}\mathcal{B}_{2}^{T}\mathcal{B}_{1}^{\ast}M_{R}& \\ \nonumber
&\Rightarrow M_{R}^{\prime} = M_{R} + \frac{1}{2}(M_{D}M_{D}^{\dagger}M_{R}^{-1} + M_{S}M_{S}^{\dagger}M_{R}^{-1} + M_{R}^{-1}M_{D}^{\ast}M_{D}^{T} + M_{R}^{-1}M_{S}^{\ast}M_{S}^{T}) & \\ \nonumber
&~~~~~~~~~~~~~~~  + \frac{1}{2}(M_{S}\mu^{\ast}M_{S}^{T}M_{R}^{-2} + M_{R}^{-2}M_{S}\mu^{\ast}M_{S}^{T})& \\ \nonumber
& \Rightarrow M_{R}^{\prime} = M_{R} + \frac{1}{2} [(M_{D}M_{D}^{\dagger} + M_{S}M_{S}^{\dagger})M_{R}^{-1} + M_{S}\mu^{\ast}M_{S}^{T}M_{R}^{-2} + {\rm Trans.}] \simeq M_{R}.&
\end{flalign}
Substituting the values of $\mathcal{B}$ from Eq.~\eqref{eq:Bvalues_int} to Eq.~\eqref{eq:U1prime} and considering expansion up to $M_{R}^{-1}$ we get,
\begin{footnotesize}
\begin{flalign}
\label{U1prime_final}
\mathcal{U}_{1}^{\prime} = \begin{pmatrix} 1 - \frac{1}{2}M_{D}^{\dagger}M_{R}^{-2}M_{D} & - \frac{1}{2}M_{D}^{\dagger}M_{R}^{-2}M_{S} & M_{D}^{\dagger}M_{R}^{-1} \cr
- \frac{1}{2}M_{S}^{\dagger}M_{R}^{-2}M_{D} & 1 - \frac{1}{2}M_{S}^{\dagger}M_{R}^{-2}M_{S} & M_{S}^{\dagger}M_{R}^{-1} + \mu^{\ast}M_{S}^{T}M_{R}^{-2} \cr
-M_{R}^{-1}M_{D} & - (M_{R}^{-1}M_{S} + M_{R}^{-2}M_{S}^{\ast}\mu) & 1- \frac{1}{2}M_{R}^{-1}(M_{D}M_{D}^{\dagger}+M_{S}M_{S}^{\dagger})M_{R}^{-1} \end{pmatrix}.
\end{flalign}
\end{footnotesize}
Using the expressions of Eqs.~\eqref{eq:mheavy_int} and \eqref{eq:mlight_int} (neglecting powers lower than $M_{R}^{-1}$) we have the intermediate block diagonalization matrix as
\begin{flalign}
\mathcal{\hat{M}}_{bd} = \begin{pmatrix} -M_{D}^{T}M_{R}^{-1}M_{D} & -M_{D}^{T}M_{R}^{-1}M_{S} & 0\cr
                                          -M_{S}^{T}M_{R}^{-1}M_{D} & \mu -M_{S}^{T}M_{R}^{-1}M_{S} & 0 \cr
                                               0 & 0 & M_{R} \end{pmatrix}.
\end{flalign}
For further block-diagonalization, we again follow the same prescriptions~\cite{Grimus:2000vj, Mitra:2011qr}. Following the similar \textit{ansatz}, we have,
\begin{equation}
\label{eq:U1doubleprime}
\mathcal{U}_{1}^{\prime\prime} = \begin{pmatrix} \sqrt{1 - \mathcal{B}^{\prime}{\mathcal{B}^{\prime}}^{\dagger}} &  \mathcal{B}^{\prime} \cr - {\mathcal{B}^{\prime}}^{\dagger} &   \sqrt{1 - {\mathcal{B}^{\prime}}^{\dagger}\mathcal{B}^{\prime}} \end{pmatrix},
\end{equation}
and from similar diagonalization 
\begin{align}
\label{eq:diagtwo}
\mathcal{U}_{1}^{\prime \prime T}\mathcal{\hat{M}}_{bd}\mathcal{U}_{1}^{\prime\prime} = \mathcal{M}_{bd} =
\begin{pmatrix} 
\mathcal{M}_\text{light} & 0 \\ 
0 & \mathcal{M}_\text{heavy} 
\end{pmatrix},
\end{align}
we obtain
\begin{flalign}
\label{eq:cond2}
&\mathcal{B}^{\prime T}(-M_{D}^{T}M_{R}^{-1}M_{D})\sqrt{1 - \mathcal{B}^{\prime}\mathcal{B}^{\prime \dagger}} - \mathcal{B}^{\prime T}(-M_{D}^{T}M_{R}^{-1}M_{S})\mathcal{B}^{\prime \dagger}& \nonumber \\ 
&+ \sqrt{1 - \mathcal{B}^{\prime T}\mathcal{B}^{\prime\ast}}(-M_{S}^{T}M_{R}^{-1}M_{D})\sqrt{1 - \mathcal{B}^{\prime}\mathcal{B}^{\prime\dagger}} - \sqrt{1 - \mathcal{B}^{\prime T}\mathcal{B}^{\prime\ast}}(\mu -M_{S}^{T}M_{R}^{-1}M_{S})  \mathcal{B}^{\prime\dagger} = 0.&
\end{flalign}
Assuming $\mathcal{B}_{j}^{\prime} \propto \frac{1}{M_{S}^{j}}$, expressions of different $\mathcal{B}^{\prime}$ can be obtained by equating different coefficients of $M_{S}^{j}$ for different values of $j$ from Eq.~\ref{eq:cond2}:
\begin{gather}
\label{eq:B1prime}
-M_{S}^{T}M_{R}^{-1}M_{D} + M_{S}^{T}M_{R}^{-1}M_{S}\mathcal{B}_{1}^{\prime\dagger}  = 0,\\
\label{eq:B2prime}
M_{S}^{T}M_{R}^{-1}M_{S}\mathcal{B}_{2}^{\prime\dagger} = 0,\\
\label{eq:B3prime}
M_{S}^{T}M_{R}^{-1}M_{S}\mathcal{B}_{3}^{\prime\dagger} = \mu \mathcal{B}_{1}^{\prime\dagger}.~({\rm ignoring~ less~ contributing~ terms})
\end{gather}
From the above equations,
\begin{align}
\label{eq:Bprimevalues}
\mathcal{B}_{1}^{\prime\dagger} = M_{S}^{-1}M_{D},~ \mathcal{B}_{2}^{\prime} = 0,~ \mathcal{B}_{3}^{\prime\dagger} = M_{S}^{-1}M_{R}{M_{S}^{T}}^{-1}\mu M_{S}^{-1}M_{D}.
\end{align}
Since $\mathcal{B}_{2}^{\prime} =0$,
\begin{footnotesize}
\begin{align}
\label{eq:mlight}
\mathcal{M}_\text{light} = &\left( 1 - \frac{1}{2}\mathcal{B}_{1}^{\prime\ast}\mathcal{B}_{1}^{\prime T}\right)(-M_{D}^{T}M_{R}^{-1}M_{D})\left(1- \frac{1}{2}\mathcal{B}_{1}^{\prime}\mathcal{B}_{1}^{\prime\dagger}\right) - \left(1 - \frac{1}{2}\mathcal{B}_{1}^{\prime\ast}\mathcal{B}_{1}^{\prime T}\right)(-M_{D}^{T}M_{R}^{-1}M_{S})(\mathcal{B}_{1}^{\prime\dagger}) \nonumber \\ 
&-\mathcal{B}_{1}^{\prime\ast}(-M_{S}^{T}M_{R}^{-1}M_{D})\left(1- \frac{1}{2}\mathcal{B}_{1}^{\prime}\mathcal{B}_{1}^{\prime\dagger}\right) + \mathcal{B}_{1}^{\prime\ast}(\mu -M_{S}^{T}M_{R}^{-1}M_{S})\mathcal{B}_{1}^{\prime\dagger} = M_{D}^{T}{M_{S}^{-1}}^{T}\mu M_{S}^{-1}M_{D},
\end{align}
\end{footnotesize}
and,
\begin{footnotesize}
\begin{align}
\label{eq:mheavy}
\mathcal{M}_\text{heavy} &= \mathcal{B}_{1}^{\prime T}(-M_{D}^{T}M_{R}^{-1}M_{D})\mathcal{B}_{1}^{\prime} + \mathcal{B}_{1}^{\prime T}(-M_{D}^{T}M_{R}^{-1}M_{S})\left(1- \frac{1}{2}\mathcal{B}_{1}^{\prime\dagger}\mathcal{B}_{1}^{\prime}\right) \nonumber \\  
&+ \left(1- \frac{1}{2}\mathcal{B}_{1}^{\prime T}\mathcal{B}_{1}^{\prime\ast}\right)(-M_{S}^{T}M_{R}^{-1}M_{D})\mathcal{B}_{1}^{\prime} + 
 \left(1- \frac{1}{2}\mathcal{B}_{1}^{\prime T}\mathcal{B}_{1}^{\prime\ast}\right)(\mu-M_{S}^{T}M_{R}^{-1}M_{S})\left(1- \frac{1}{2}\mathcal{B}_{1}^{\prime\dagger}\mathcal{B}_{1}^{\prime}\right) \nonumber \\ 
&= (\mu-M_{S}^{T}M_{R}^{-1}M_{S}) + \frac{1}{2}[M_{S}^{T}M_{R}^{-1}M_{D}M_{D}^{\dagger}{M_{S}^{-1}}^{\dagger} + \mu M_{S}^{-1}M_{D}M_{D}^{\dagger}{M_{S}^{-1}}^{\dagger} + {\rm Trans.}].
\end{align}
\end{footnotesize}
The final block-diagonalized matrix is given by
\begin{flalign}
\label{eq:finalMbd}
\mathcal{M}_{bd} = \begin{pmatrix} M_{D}^{T}{M_{S}^{-1}}^{T}\mu M_{S}^{-1}M_{D} & 0 & 0\cr
                                          0 & \mu -M_{S}^{T}M_{R}^{-1}M_{S} & 0 \cr
                                               0 & 0 & M_{R} \end{pmatrix}.
\end{flalign}
From Eqs.~\eqref{eq:U1doubleprime} and \eqref{eq:Bprimevalues} we have,
\begin{flalign}
\label{U1doubleprime_final}
\mathcal{U}_{1}^{\prime\prime} = \begin{pmatrix} 1 - \frac{1}{2}M_{D}^{\dagger}{M_{S}^{-1}}^{\dagger}M_{S}^{-1}M_{D} & M_{D}^{\dagger}{M_{S}^{-1}}^{\dagger} & 0 \cr
-M_{S}^{-1}M_{D} -M_{S}^{-1}M_{R}{M_{S}^{T}}^{-1}\mu M_{S}^{-1}M_{D} & 1 - \frac{1}{2}M_{S}^{-1}M_{D}M_{D}^{\dagger}{M_{S}^{-1}}^{\dagger} & 0 \cr
0 & 0 & 1 \end{pmatrix}.
\end{flalign}
The final block-diagonalization matrix $\mathcal{U}_{1}$ is given by
\begin{footnotesize}
\begin{align}
\label{eq:U1}
\mathcal{U}_{1} &= \mathcal{U}_{1}^{\prime} \mathcal{U}_{1}^{\prime \prime} \nonumber \\
&= \begin{pmatrix} 1 - \frac{1}{2}M_{D}^{\dagger}{M_{S}^{-1}}^{\dagger}M_{S}^{-1}M_{D} & M_{D}^{\dagger}{M_{S}^{-1}}^{\dagger} & M_{D}^{\dagger}M_{R}^{-1} \cr
-M_{S}^{-1}M_{D} & 1 - \frac{1}{2}M_{S}^{-1}M_{D}M_{D}^{\dagger}{M_{S}^{-1}}^{\dagger} - \frac{1}{2}M_{S}^{\dagger}M_{R}^{-2}M_{S} & M_{S}^{\dagger}M_{R}^{-1} \cr
{M_{S}^{T}}^{-1}\mu M_{S}^{-1}M_{D} & - M_{R}^{-1}M_{S} & 1- \frac{1}{2}M_{R}^{-1}M_{S}M_{S}^{\dagger}M_{R}^{-1} \end{pmatrix}.
\end{align}
\end{footnotesize}
Finally to the leading order the active and heavy neutrino mass matrices are given by
\begin{eqnarray}
m_{\nu} &\sim& M_{D}^{T}{M_{S}^{-1}}^{T}\mu M_{S}^{-1}M_{D},\\
m_{s} &\sim& \mu -M_{S}^{T}M_{R}^{-1}M_{S}, \\
m_{n} &\sim& M_{R}.
\end{eqnarray}
In the above, if $\mu$ is greater than $M_{S}^{T}M_{R}^{-1}M_{S}$ then $m_{s}$ will be defined mainly by $\mu$ and in the opposite case, it will be dominated by $M_{S}^{T}M_{R}^{-1}M_{S}$. If $\mu \sim M_{S}^{T}M_{R}^{-1}M_{S}$ then we have to consider the next order contribution from Eq.~\eqref{eq:mheavy}.
%
%
The active and sterile matrices will be diagonalized as
\begin{eqnarray}
\label{eq:diangonalizationU2}
U^{T}m_{\nu}U &=& \text{diag}(m_{\nu_{i}}) \\
W_{s}^{T}m_{s}W_{s} &=& \text{diag}(m_{s_{i}}) \\
W_{N}^{T}m_{n}W_{N} &=& \text{diag}(M_{R_{i}}).
\end{eqnarray}

So, the mixing matrix for above block diagonalization can be written as
\begin{equation}
\label{eq:U2}
\mathcal{U}_{2} = \begin{pmatrix} U & 0 & 0 \cr 0 & W_{s} & 0 \cr 0 & 0 & W_{N} \end{pmatrix}.
\end{equation}

Finally the diagonalization matrix of Eq.~\eqref{eq:extcsawMat} can be written as,
\begin{scriptsize}
\begin{align}
\label{eq:Ufinal}
\mathcal{U} = 
\begin{pmatrix} (1 - \frac{1}{2}M_{D}^{\dagger}{M_{S}^{-1}}^{\dagger}M_{S}^{-1}M_{D})U & M_{D}^{\dagger}{M_{S}^{-1}}^{\dagger}W_{s} & M_{D}^{\dagger}M_{R}^{-1}W_{N} \cr
-M_{S}^{-1}M_{D}U & (1 - \frac{1}{2}M_{S}^{-1}M_{D}M_{D}^{\dagger}{M_{S}^{-1}}^{\dagger} - \frac{1}{2}M_{S}^{\dagger}M_{R}^{-2}M_{S})W_{s} & M_{S}^{\dagger}M_{R}^{-1}W_{N} \cr
{M_{S}^{T}}^{-1}\mu M_{S}^{-1}M_{D}U & - M_{R}^{-1}M_{S}W_{s} & (1- \frac{1}{2}M_{R}^{-1}M_{S}M_{S}^{\dagger}M_{R}^{-1})W_{N} \end{pmatrix}.
\end{align}
\end{scriptsize}

\bigskip

\bibliographystyle{JHEP}
\bibliography{extseesaw}
\end{document}